\documentclass[12pt]{article}
\usepackage{epsf}

\usepackage{epsfig,graphics}
\usepackage {graphicx}
\usepackage {epsfig}
\usepackage{subcaption}
\usepackage {tabularx} 
\usepackage{rotate}	
\usepackage{slashed}
\usepackage{bbm}
\usepackage{color}
\usepackage{tikz}
\usetikzlibrary{decorations.pathmorphing} 
\usepackage{amsmath}
\usepackage{amsfonts}
\usepackage{amssymb}
\usepackage{graphicx}
\usepackage{cite}

\usepackage{fancyhdr}
\usepackage{hyperref}
\usepackage{diagbox}

\newcommand{\bmat}{\left(\begin{array}}
\newcommand{\emat}{\end{array}\right)}

\def\yzero{\smash{\hbox{$y\kern-4pt\raise1pt\hbox{${}^\circ$}$}}}
\def\p{\partial}
\def\a{\alpha}
\def\b{\beta}

\def\beq{\begin{equation}}
\def\eeq{\end{equation}}
\def\beqa{\begin{eqnarray}}
\def\eeqa{\end{eqnarray}}
\def\Om{\Omega}

\def\-{\hphantom{-}}

\def\s2{\frac{1}{\sqrt2}}

\def\oh{\frac{1}{2}}

\def\ch{{\cal H}}

\def\Dsl{\,\raise.15ex\hbox{/}\mkern-13.5mu D} 
\def\IZ{Z\kern-.4em  Z}

\def\CK {{\cal K}}

\def\be{\begin{equation}}
\def\ee{\end{equation}}
\def\bea{\begin{eqnarray}}
\def\eea{\end{eqnarray}}
\def\bes{\begin{subequations}}
\def\ees{\end{subequations}}

\def\oh{\frac{1}{2}}

\def\re{\mbox{Re}\, }
\def\im{\mbox{Im}\, }

\def\Z{\mathbbm{Z}}

\catcode`\@=11   
\newdimen\@rotdimen
\newbox\@rotbox  

\def\@vspec#1{\special{ps:#1}}
\def\@rotstart#1{\@vspec{gsave currentpoint currentpoint translate
   #1 neg exch neg exch translate}}
\def\@rotfinish{\@vspec{currentpoint grestore moveto}}
\def\@rotr#1{\@rotdimen=\ht#1\advance\@rotdimen by\dp#1%
   \hbox to\@rotdimen{\hskip\ht#1\vbox to\wd#1{\@rotstart{90 rotate}%
   \box#1\vss}\hss}\@rotfinish}
\def\@rotl#1{\@rotdimen=\ht#1\advance\@rotdimen by\dp#1%
   \hbox to\@rotdimen{\vbox to\wd#1{\vskip\wd#1\@rotstart{270 rotate}%
   \box#1\vss}\hss}\@rotfinish}%
\def\@rotu#1{\@rotdimen=\ht#1\advance\@rotdimen by\dp#1%
   \hbox to\wd#1{\hskip\wd#1\vbox to\@rotdimen{\vskip\@rotdimen
   \@rotstart{-1 dup scale}\box#1\vss}\hss}\@rotfinish}%
\def\@rotf#1{\hbox to\wd#1{\hskip\wd#1\@rotstart{-1 1 scale}%
   \box#1\hss}\@rotfinish}%
\def\rotate{\@ifnextchar[{\@rotate}{\@rotate[l]}}
\def\@rotate[#1]#2{\setbox\@rotbox=\hbox{#2}\@nameuse{@rot#1}\@rotbox}

\catcode`\@=12

\topmargin
-1.5cm
\textwidth
15.5cm
\textheight
23.5cm
\oddsidemargin
0.7cm
\evensidemargin
0.7cm

\begin{document}

\makeatletter
\@addtoreset{equation}{section}
\makeatother
\renewcommand{\theequation}{\thesection.\arabic{equation}}
\pagestyle{empty}
\vspace{-0.2cm}
\rightline{ IFT-UAM/CSIC-24-92}
\vspace{0.5cm}
\begin{center}



\LARGE{On  small  Dirac Neutrino Masses \\  in String Theory}
\\[8mm]
\large{Gonzalo F.~Casas,$^\diamondsuit$  Luis E.~Ib\'a\~nez$^{\clubsuit  \diamondsuit}$ and Fernando  Marchesano$^\diamondsuit$}
\\[6mm]
\small{$^\clubsuit$  Departamento de F\'{\i}sica Te\'orica \\ Universidad Aut\'onoma de Madrid,
Cantoblanco, 28049 Madrid, Spain}  \\[5pt]
$^\diamondsuit$  {Instituto de F\'{\i}sica Te\'orica UAM-CSIC, c/ Nicolas Cabrera 13-15, 28049 Madrid, Spain} 
\\[6mm]
\small{\bf Abstract} \\ [6mm]
\end{center}
\begin{center}
\begin{minipage}[h]{15.22cm}

We study how tiny Dirac neutrino masses consistent with experimental constraints can arise in string theory SM-like vacua. We use as a laboratory 4d ${\cal N}=1$ type IIA Calabi--Yau orientifold compactifications, and in particular recent results on Yukawa couplings at infinite field-space distance. In this regime we find Dirac neutrino masses of the form $m_\nu \simeq  g_\nu\langle H\rangle$, with $g_{\nu}$ the gauge coupling of the massive $U(1)$ under which the right-handed neutrinos $\nu_R$ are charged, and which should be in the range $g_{\nu}\simeq 10^{-14}-10^{-12}$ to reproduce neutrino data. The neutrino mass suppression occurs because the right-handed neutrino kinetic term behaves as $K_{\nu\nu} \simeq 1 /g_{\nu}^2 $. At the same time a tower of $\nu_R$-like  states appears with characteristic scale $m_0\simeq g_{\nu}^2M_{\rm P}\simeq 0.1-500$ eV, in agreement with Swampland expectations.
Two large hidden dimensions only felt by the $\nu_R$  sector arise at the same scale, while the string scale is around $M_s\simeq g_\nu M_{\rm P}\simeq 10-700$ TeV. Some phenomenological implications and model building challenges are described.
We also describe the difficulties in obtaining appropriate tiny neutrino Yukawas in the cases of a single or more than two large dimensions. Thus the case with two large dimensions seems to be quite unique.

\end{minipage}
\end{center}
\newpage
\setcounter{page}{1}
\pagestyle{plain}
\renewcommand{\thefootnote}{\arabic{footnote}}
\setcounter{footnote}{0}



\tableofcontents

 %
 	
\section{Introduction}
\label{s:intro}
 Neutrinos remain possibly the most mysterious particles of the Standard Model (SM)
 (for a recent review  see e.g. \cite{SajjadAthar:2021prg,Esteban:2020cvm} and references therein). We still do not know whether they are Dirac or Majorana particles and what is the origin of their tiny mass. In fact, we still cannot rule out one of the three neutrino types being massless, and we ignore whether there is a normal hierarchy (NH) or an inverted hierarchy (IH) in their mass pattern. Depending on the origin of their mass, two main possibilities arise:
 \begin{itemize}
 \item {\it Dirac masses}.
 In this case it  is assumed that
 there are both left-handed $\nu_L$  (in a $SU(2)_L$ doublet) and right-handed $\nu_R$
 (SM singlet) neutrinos, both initially massless.  They  couple  to the Higgs boson  through  Yukawa couplings $Y_\nu$ and 
 once the Higgs gets a vev they get Dirac masses given by
 \be
 m_{\nu}^{\rm Dirac} \, \simeq \,  Y_\nu \langle H_0 \rangle \, .
 \ee
 Here $Y_\nu$ are Yukawa couplings which should be tiny, in the range $10^{-12} - 10^{-13}$, in order to match the experimental
 oscillation data, and $\langle H_0 \rangle$ is the Higgs vev.
 
  \item {\it Majorana masses}.
  One may consider such tiny Yukawa couplings to be fine-tuned, which would be nice to avoid. 
 A natural way in which tiny Yukawa couplings are avoided  is the celebrated `see-saw mechanism'
 \cite{Minkowski:1977sc,Gell-Mann:1979vob,Yanagida:1980ph}. 
 Since the $\nu_R$'s are  SM singlets they may be arbitrarily heavy,  perhaps with masses $M_R$ of order some UV scale,
 so that $M_R\gg M_{EW}$. This structure leads to the well-known expression for
  Majorana masses of order
 \be
 m_\nu^{\rm Majorana} \, \simeq \, 
 \frac { \left(Y_\nu \langle H_0\rangle\right)^2} {M_R}\, ,
 \ee
 which for ordinary Yukawa couplings of order $10^{-4} - 10^{-1}$ will give masses consistent with experiment for  $M_R\sim 10^{14}$ GeV. These eigenstates are predominantly coming from the $\nu_L$'s and have Majorana masses.
 
 \end{itemize}
 
 It is interesting to explore how, if at all, these two mechanisms arise in string theory. A fruitful strategy to explore these question has been to consider string compactifications based in type IIA CY orientifolds with intersecting D6-branes (see e.g. \cite{Blumenhagen:2005mu,Blumenhagen:2006ci,Marchesano:2007de,Lust:2009kp,Ibanez:2012zz,Marchesano:2022qbx,Marchesano:2024gul}
 and references therein), 
 and this is the class of theories that we will use in this paper as a laboratory.
 Due to different dualities transforming this class of vacua to other corners in the 4d ${\cal N}=1$ landscape, we believe that the results we obtain are rather generic in the space of realistic string vacua. 
 
 \vspace*{.3cm}
 
 {\bf  Large $\nu_R$ Majorana masses in string theory}

  \vspace*{.2cm}

 The simplest setup assumes type  IIA  CY orientifold vacua close to the SM physics, with $3+2+n$ stacks of intersecting D6-branes giving rise to a  $U(3)\times U(2)\times U(1)^n$, $n\geq 2$ gauge group. The subgroup $U(1)^{n+2}$ contains the hypercharge and the rest of the $U(1)$'s become generically massive from $B\wedge F$ couplings, inducing St\"ukelberg masses for them.  Here $B$ are 2-forms arising from the gravitational sector of the theory.  The extra, massive $U(1)$'s include lepton and baryon numbers which are thus gauged. The spectrum always includes perturbatively massless right-handed neutrinos $\nu_R$. However, it was shown 
\cite{Ibanez:2006da,Blumenhagen:2006xt,Cvetic:2007ku} that  once the  lepton number   gets massive due to a $B \wedge F$ coupling, $\nu_R$ Majorana masses may be induced by `charged stringy instantons'. These correspond to Euclidean $D2$-branes wrapping 3-cycles in the internal manifold and intersecting the branes that localise the $\nu_R$'s \cite{Ibanez:2006da,Blumenhagen:2006xt,Cvetic:2007ku}. The induced masses are of order
 \be
 m_{\nu_R}\, \simeq \, M_s\, e^{-S_E}  \,  \simeq \, M_s\, e^{-u_E} \ ,
 \ee
 where $M_s$ is the string scale and $S_E$ is the instanton action, whose size is determined by a complex structure scalar $u_E$  that measures the volume of the 3-cycle wrapped by the instanton $E2$ in string units. If this volume is not large and we have a high string scale around $M_s\simeq 10^{16}$ GeV, one may obtain large Majorana masses for the right-handed neutrinos in the range consistent with experiment.  This is an elegant mechanism.  Still it turns out that the conditions for such instantons to produce a mass term in specific string vacua are not so easy to attain. In particular, it turns out that typically there are additional instanton  `neutral zero modes' in the theory which often lead to vanishing amplitudes for a mass term. In particular,  in \cite{Ibanez:2007rs} a search was made in a large class of type II 4d ${\cal N}=1$  SM-like orientifolds obtained from rational CFT's (Gepner-like  models). It was found no example in which the required structure of zero modes to get $\nu_R$ Majorana masses exists. This should not be considered as a sort of `no-go theorem' since the class of theories studied was large but certainly not exhaustive. Furthermore additional effects like closed string fluxes may get rid of some unwanted singlet zero modes, allowing for masses to be generated. Nevertheless, this motivates to study whether the Dirac mass alternative is viable in string theory.

 \vspace*{.3cm}
 
 {\bf Tiny Dirac neutrino masses in string theory}

  \vspace*{.2cm}
 
 This is the alternative that we study in the present paper. Note that within a given string vacuum three  conditions are  then required:
 1) There must exist some limit in moduli space in which tiny Dirac masses for all neutrinos are obtained. This limit must be such that 2) the Yukawa couplings as well as the SM gauge couplings remain \textit{unsuppressed} for the rest of the fermions, quarks and charged leptons, while 3) right-handed Majorana neutrino masses are very small or vanishing. It turns out that this third condition is essentially automatic, once the 
 other two are achieved. On the contrary,  we find that the two first conditions are very strong and almost uniquely fix how the
 structure of mass scales must be in a string vacuum in order to get viable Dirac neutrino masses.
 The subject of the first condition has been recently addressed in the paper 
 \cite{Casas:2024ttx} in which it has been studied 
 the behaviour of matter Yukawa couplings at large moduli, leading to vanishing small Yukawa couplings. There it was found that
 the Yukawa couplings among three chiral matter fields living at  D6-brane intersections $i,j,k$ have the general structure

 \be
	 Y_{ijk}\, =\, e^{\phi_4/2} {\rm Vol}_X^{1/4} W_{ijk}  \Theta_{ijk}^{1/4} \, .
	 \ee
 On the one hand, ${\rm Vol}_X$ is the volume of the compact manifold in units of $M_s$ and $W_{ijk}$ is the holomorphic superpotential, both depending only
 on the complexified K\"ahler moduli $T^a$.  On the other hand $\phi_4$ is the 4d dilaton and the functions  $\Theta_{ijk}$ encode the information about the intersection angles of the three D6-branes involved in the Yukawa coupling, both quantities depending only on the complex structure fields $u^K$. As pointed out in \cite{Casas:2024ttx}, the interesting limit for us will be the one in which we keep the K\"ahler moduli fixed and ${\rm Vol}_X$ is not too large. In that case, limits with vanishing Yukawas imply some complex structure fields becoming large. In a limit parametrised by a single complex structure direction $u\rightarrow \infty$,  Yukawa couplings behave like $Y\simeq 1/u^r$ with  $r$ some rational number typically in the range $1/4 \leq r\leq 1$ \cite{Casas:2024ttx}. At the same time towers of light states with the same quantum numbers as the massless quarks and leptons, termed `gonions' in \cite{Aldazabal:2000cn}, become light.
  
   In the case of neutrinos, the crucial issue is to obtain tiny neutrino Yukawa couplings while the rest of Yukawa couplings as well as the SM gauge couplings, remain unsuppressed.  We find that in the context of type II CY SM-like orientifolds, tiny Yukawa couplings require the presence of $\nu_R$ gonion towers becoming light as $m_{{\rm gon},\nu_R}\sim M_{\rm P}/u$ while two dimensions become large at the same rate. The appearance of precisely two large dimensions is a consequence of  requiring that {\it i)} only neutrinos get very suppressed Yukawa couplings and {\it ii)} the string scale $M_s$ is compatible with experiment. This is essentially because all the relevant scales of this scenario have simple behaviour in terms of $u$. In particular, the neutrino Yukawa couplings behave like $Y_\nu\sim u^{-1/2}$, which is also the rate of variation $g_\nu \sim u^{-1/2}$ of the gauge coupling associated with the massive $U(1)_\nu$ gauge interaction felt by right-handed neutrinos. 
  The fundamental scales of the theory are then  determined in terms of the neutrino couplings as
  \beqa
  M_s &  \simeq &Y_\nu M_{\rm P} \, \simeq \, g_\nu M_{\rm P} \,  , \\
     m_{{\rm gon},\nu_R}  & \simeq &  m_{{\rm KK}} \, \simeq \, Y_\nu^2\, M_{\rm P}\, \simeq \, g_\nu^2 \, M_{\rm P} \, ,
  \eeqa
 where $Y_\nu$ here is the largest of the neutrino Yukawa couplings. Setting $Y_\nu\simeq 7\times 10^{-13}$, consistent with 
 neutrino oscillations data, gives
 \be
 M_s \simeq 700\ {\rm TeV} \,  , \qquad  m_{{\rm gon},\nu_R}  \, \simeq  \,  m_{{\rm KK}} \, \simeq \, 500 \, {\rm eV}\, .
 \ee
  As a result the string scale is lowered to a scale well above, but not far away, from the EW scale. In addition there are two
  large dimensions around the 500 eV, along with a tower of $\nu_R$-like states. There is also an extra $U(1)_\nu$  with very weak coupling $g_\nu\sim 10^{-14} -10^{-12}$  with a mass $M_{V_\nu}$ in the wide range  $0.1 \, {\rm eV}\leq M_{V_\nu} \leq M_s$. In fact, if the  $\nu_R$ gonion towers are not universal, the scales may be lowered down to the values 
 \be
 M_s \simeq 10\ {\rm TeV} \, ,  \qquad  m_{{\rm gon},\nu_R}  \, \simeq  \,  m_{{\rm KK}} \, \simeq \,  0.1 \, {\rm eV} \ ,
 \ee
not far above LHC bounds for the string scale:  $M_s\lesssim 8 $ TeV. 
  We briefly discuss possible phenomenological implications of the above spectrum of
  new particles in section \ref{phenoresults}.
The obtained structure may be considered as a concrete string theory realisation of the
 Large Extra Dimension (LED) scenario of \cite{Arkani-Hamed:1998sfv} and neutrino masses in this setting \cite{Dvali:1999cn}. However, as seen above, the embedding into a string theory lead us to a very concrete scenario in which two dimensions must be large and the relevant mass scales display a very specific structure.

  The results obtained for the mass scales and towers of states in these types of compactifications are, as expected, consistent with general Swampland arguments (see \cite{Brennan:2017rbf,Palti:2019pca,vanBeest:2021lhn,Grana:2021zvf} for reviews). In particular light towers of states are obtained along large complex structure  limits, in agreement with the Swampland Distance Conjecture (SDC),
  and  such towers obey the Weak Gravity Conjecture  (WGC) for the $U(1)_\nu$ gauge boson coupling to $\nu_R$'s. In fact there are arguments, based on the  AdS instability conjecture \cite{Ooguri:2016pdq,Freivogel:2016qwc} which state that neutrinos must be Dirac and the lightest neutrino must have a mass bounded by the cosmological constant scale,
  $m_{\nu}^{\rm min} \lesssim \Lambda_{{\rm cc}}^{1/2}$ \cite{Ibanez:2017kvh,Hamada:2017yji,Gonzalo:2021zsp}.  This nicely fits with the Dirac option which we analyse in this paper. We describe some
  consequences of these Swampland arguments in section \ref{s:thecc}. In combination with our previous results, they give a rationale for why two large dimensions should open:  they are required to obtain a sufficiently light Dirac neutrino to obey the Swampland bound  $m_{\nu}^{\rm min} \lesssim \Lambda_{{\rm cc}}^{1/4}$. This also explains the apparent coincidence between the neutrino mass scale and that of the cosmological constant.
  
  The structure of the rest of the paper is as follows. In section \ref{s:orientifolds}, we give a short introduction to type IIA ${\cal N}=1$ CY orientifolds with intersecting D6-branes. We describe the spectrum in this class of theories and the structure of  asymptotically  vanishing Yukawa couplings. We also provide an explicit  example that illustrates the general limiting structure of mass scales and  Yukawas. In section \ref{s: SM intersecting} we give a D6-brane configuration whose intersections contain the massless chiral fields of the SUSY SM and Higgs fields.  We discuss how tiny Yukawa couplings for neutrinos may be obtained in a setting with two large dimensions, and describe the general structure of scales and the mass of the extra $U(1)$'s. In section \ref{phenoresults} we impose that the obtained value for the neutrino Yukawa couplings agrees with neutrino data,  which essentially fixes all scales. We also give a brief description of possible phenomenological implications.   In section \ref{s:thecc} we briefly describe how the results obtained fit within general Swampland arguments and in particular the implications of the upper bound on the lightest neutrino mass coming from the 3d SM compactification and
  the AdS instability conjecture. Some final comments are left for section \ref{s:conclusions}.
 
  Some technical details have been relegated to three appendices.  In appendix \ref{onedimension} we explain the difficulties of having a SM-like orientifold with a single large dimension and neutrino Yukawa couplings small, and then rule out settings with more than two large dimensions because they imply a very small string scale.  In appendix  \ref{su5} we present an $SU(5)$ intersection brane setting in which the same general structure for obtaining tiny Yukawa couplings is realised. Appendix \ref{anomalias} contains an analysis of the $U(1)$ anomalies and the conditions for  $U(1)$'s other than hypercharge becoming massive in the explicit SM-like configuration described in the main text.

	\section {4d \texorpdfstring{${\cal N}=1$}{Lg} type IIA orientifolds as a laboratory }\label{s:orientifolds}

 In this section we first briefly review the construction of type IIA ${\cal N}=1$ CY orientifolds with intersecting
 D6-branes. We then describe the structure of matter field Yukawa couplings and their behavior for large moduli,
 leading to the search for small Yukawa couplings. The main characteristics of these constructions 
 at infinite distance are illustrated in a $\mathbbm{Z}_2\times \mathbbm{Z}_2$ toy model with a Pati-Salam structure. 
 We also describe the difficulties in getting a compactification with only one large dimension if we want to keep unsuppressed the SM gauge and Yukawa couplings other than neutrino's.

	\subsection{Type IIA orientifolds and intersecting D6-branes}

Let us sketch the basic properties of the string theory construction that we will use to analyse Yukawa couplings, namely type IIA Calabi--Yau (CY) orientifolds with intersecting D6-branes. A  detailed description of this class of vacua and their phenomenology can be found in the reviews \cite{Blumenhagen:2005mu,Blumenhagen:2006ci,Marchesano:2007de,Lust:2009kp,Ibanez:2012zz,Marchesano:2022qbx,Marchesano:2024gul} as well as in \cite[section 2.1]{Casas:2024ttx}, whose conventions and notation we will follow. 

Calabi--Yau orientifold vacua are built on a background of the form $X_4 \times X_6$, where $X_4$ are four macroscopic dimensions and $X_6$ are six extra dimensions curled up in a compact manifold with a CY metric. One then imposes a parity-reversal $\mathbbm{Z}_2$ quotient to the worldsheet of strings propagating in this background. In terms of 10d geometry, this translates into a quotient by an involution ${\cal R}$ acting on $X_6$ and leaving $X_4$ invariant. In type IIA CY orientifolds, this action is given by ${\cal R} (J, \Om) = (-J, \bar{\Om})$, where $J$ is the K\"ahler form and  $\Omega$ the holomorphic three-form that determine the CY metric on $X_6$. The fixed point set of ${\cal R}$ is of the form $X_4 \times \Pi_{\rm O6}$, with $\Pi_{\rm O6}$ either one or a sum of smooth three-cycles, where the O6-plane content of the compactification is located. To achieve a consistent background one needs to cancel the internal charge induced by the O6-planes, for instance by adding stacks of $N_\a$ D6-branes on $X_4 \times \Pi_\a$, where $\Pi_\a$ are a set of three-cycles such that $\sum_\a N_\a \left( [\Pi_\a] + [\Pi_{\a*}] \right) = 4 [\Pi_{\rm O6}]$.\footnote{Other possibilities to cancel tadpoles are adding background fluxes or coisotropic D8-branes \cite{Font:2006na}.} Here the brackets denote their  homology classes in $H_3(X_6, \Z)$ and  $\Pi_{\a*} = {\cal R} (\Pi_\a)$ stands for the orientifold image of each D6-brane. To insure instability, one typically requires a BPS condition for the $\Pi_\a$, namely that they are special Lagrangian three-cycles with no worldvolume flux. 

The 4d EFT obtained from this compactification has an open and a closed string sector, each of them hosting a  
set of light fields. Closed strings give rise to the 4d ${\cal N}=1$ gravity multiplet, as well as a set of chiral multiplets whose bosonic components describe the internal geometry of $X_6$. These are the K\"ahler $T^a = b^a + i t^a$ and complex structure $U^K = \zeta^K + i u^K$ moduli. The real part of these fields are periodic axionic variables, while the imaginary part are saxion fields that describe the geometry of $X_6$ in terms of $J$ and $\re \Omega$, respectively. They also determine the kinetic terms of the closed string sector, via the K\"ahler potential $K \equiv K_K + K_Q$, where \cite{Grimm:2004ua}
\be
K_K   \equiv   -{\rm log} \left({\rm Vol}_{X_6}\right) = -{\rm log} \left(\frac{i}{48} \CK_{abc} (T^a - \bar{T}^a)(T^b - \bar{T}^b)(T^c - \bar{T}^c) \right) \, ,
\label{KK}
\ee
with ${\cal K}_{abc} \equiv \ell_s^{-6} \int_{X_6} \omega_a \wedge \omega_b \wedge \omega_c$ the triple intersection numbers of $X_6$ and 
\begin{equation}
 K_Q \equiv -2 \log \ch = - 2 \log \left( \frac{i}{8\ell_s^6} \int_{X_6} e^{-2\phi}  \Om \wedge \bar{\Om} \right)  = 4\phi_4 \, ,
 \label{KQ}
\end{equation}
with $\phi_4 = \phi - \oh \log {\rm Vol}_{X_6}$ the 4d dilaton and $\phi$ the 10d dilaton. 

\setlength{\belowcaptionskip}{0pt}
\begin{table}[htb]
\renewcommand{\arraystretch}{1.25}
\begin{tabular}{ll}
\hline
Non-Abelian gauge group & $\prod_\a SU(N_\a)$\\
Massless $U(1)$s &  $\sum_\a c_\a U(1)_\a$ such that $\sum_\a c_\a ([\Pi_{\a}] - [\Pi_{\a*}]) = 0$  \\
Chiral multiplets & $\sum_{\a<\b}\, I_{\a\b} ({\bf N}_\a, {\bf \bar{N}}_\b) \, + \, I_{\a\b*} ({\bf N}_\a, {\bf \bar{N}}_\b) $\\
\hline
\end{tabular}
\caption{4d EFT chiral spectrum, in terms of the intersection number $I_{\a\b} = [\Pi_\a] \cdot [\Pi_\b]$.  For simplicity we assume that $I_{\a\a*} = I_{\a {\rm O6}} =0$, as in the models of section \ref{s: SM intersecting}. If $\Pi_\a = \Pi_{\a*}$, the D6-brane hosts either an $SO(2N_\a)$ or $USp(2N_\a)$ gauge group.}
\label{t:specori}
\end{table}
 
The massless open string states oscillations give rise the gauge sector of the 4d EFT, which is localised at the D6-brane worldvolume and their intersections. More precisely, the net 4d chiral spectrum is specified by the topological intersection number $I_{\a\b} = [\Pi_\a] \cdot [\Pi_\b]$ between pairs of three-cycles wrapped by D6-branes, as summarised in table \ref{t:specori}. Each transverse intersection between two three-cycles localises a 4d chiral fermion at a point in $X_6$, together with an infinite tower of massive particles with the same bifundamental charges  \cite{Berkooz:1996km,Aldazabal:2000dg}. These massive states were dubbed {\em gonions} in \cite{Aldazabal:2000cn}, and they are particularly relevant in models with very small
intersection angles \footnote{ For studies of properties, couplings and production 
of these states see e.g. \cite{Anastasopoulos:2011hj,Hamada:2012wj,Anastasopoulos:2021bor,Anastasopoulos:2016yjs,Anastasopoulos:2014lpa} .}.

\begin{figure}[h]
    \centering
    \hspace{4.5em}
\begin{tikzpicture}[scale=1.250]
\draw[thick] (0,0) -- (4,0);
\draw[dashed] (-0.5,0) -- (4.5,0) node[right]{$m_{\Phi_{\a\b}}=0$};
\draw[dashed] (-0.5,2) -- (4.5,2)node[right]{$m_{{\rm gon},\a\b}=|\theta_{\a\b}|^{\frac{1}{2}}\,M_s$};
\draw[dashed] (-0.5,3.6) -- (4.5,3.6)node[right]{$M_s$};
\draw[thick] (0,2) -- (4,2);
\draw[thick] (0,2) -- (4,2);
\draw[thick] (0,2.2) -- (4,2.2);
\draw[thick] (0,2.4) -- (4,2.4);
\draw[thick] (0,2.6) -- (4,2.6);
\draw[thick] (0,2.8) -- (4,2.8);
\draw[thick] (0,3) -- (4,3);
\draw[thick] (0,3.2) -- (4,3.2);
\draw[thick] (0,3.4) -- (4,3.4);
\draw[thick] (0,3.6) -- (4,3.6);
\draw[thick] (0,3.65) -- (4,3.65);
\draw[thick] (0,3.7) -- (4,3.7);
\draw[thick] (0,3.75) -- (4,3.75);
\draw[thick] (0,3.8) -- (4,3.8);
\draw[thick,black!90] (0,3.85) -- (4,3.85);
\draw[thick,black!80] (0,3.9) -- (4,3.9);
\draw[thick,black!70] (0,3.95) -- (4,3.95);
\draw[thick,black!60] (0,4) -- (4,4);
\draw[thick,black!50] (0,4.05) -- (4,4.05);
\draw[thick,black!40] (0,4.10) -- (4,4.10);
\draw[thick,black!30] (0,4.15) -- (4,4.15);
\draw[thick,black!20] (0,4.2) -- (4,4.2);
\draw[thick,black!20] (0,4.25) -- (4,4.25);
\draw[thick,black!20] (0,4.3) -- (4,4.3);
\draw[dashed] (-0.5,4.2) -- (4.5,4.2) node[right]{\hspace{-0.2em}$M_{\rm P}$};

\draw[->,thick] (-1,0) -- (-1,4.4);
\end{tikzpicture}
\caption{Gonion tower spectrum with one small angle $\theta_{\a\b}$.
There is a tower of gonions with the same charge as the massless mode $\Phi_{\a\b}$ and
characteristic mass $m_{{\rm gon},\a\b}$. The tower runs up to the string scale $M_s$, at which a Hagedorn spectrum  is recovered. 
\label{fig: gonionspectrum}}
\end{figure}
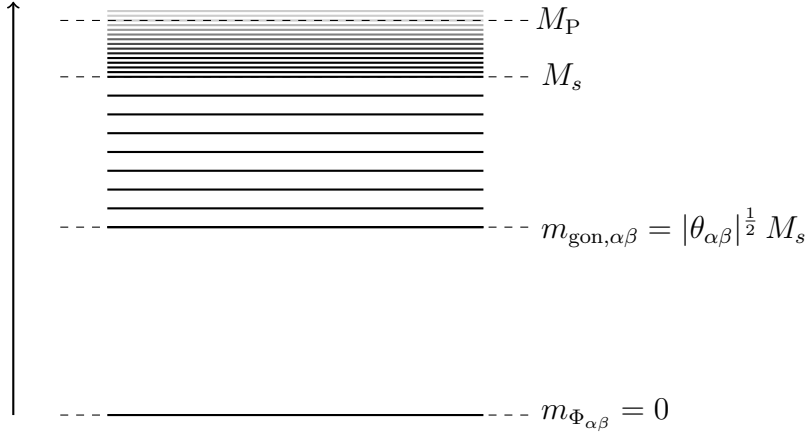

The gonion tower spectrum can be different at each intersection, depending on the local geometry. In the generic case of a transverse intersection, it is specified by three intersection angles $\theta_{\a\b}^r$ $r=1,2,3$, measured in units of $\pi$.\footnote{ Intersections specified  only in terms of two angles, dubbed ${\cal N} =2$ sectors in \cite{Casas:2024ttx},   also exist and play an important role in the construction of realistic models. See below. \label{ft:N=2}}  
Whenever one or more angles are small the tower displays a spectrum of the form \cite{Berkooz:1996km,Aldazabal:2000dg}
\be
m_{\a\b}^2 = \sum_q  k_q |\theta_{ab}^q| M_s^2\, , 
\ee
up to an angle-dependent function that depends on whether we are talking about a charged scalar, fermion or W-boson. Here  $q$ runs over those angles that satisfy $|\theta_{ab}^q| \ll 1$, $k_q \in \mathbbm{N}$ and $M_s = e^{\phi_4} M_{\rm P}$ is the  string scale. We will be mainly interested in setups with  a single small angle, in which case this tower is as dense as a two-dimensional lattice of KK modes. At the string scale, additional states appear, so that one recovers a Hagedorn-like spectrum, see figure \ref{fig: gonionspectrum} and \cite[Appendix A]{Casas:2024ttx}.

\vspace{1em}

The interplay between open and closed string sectors manifests in a set of gauge kinetic functions that depend linearly on the complex structure moduli $U^K$, and a generalised Green--Schwarz (GS) mechanism that cancels $U(1)$ anomalies and implements a St\"uckelberg mechanism on them \cite{Aldazabal:2000dg,Ibanez:2001nd}. The explicit expressions are obtained from expanding the three-cycles homology classes wrapped by the D6-branes as $[\Pi_\a] = P_{\a\, J} [\Sigma^J_+] + Q_\a^K  [\Sigma_K^-]$, where $[\Sigma^J_+]$ is a certain basis of  even and $[\Sigma_K^-]$ of odd classes under the action of ${\cal R}$, satisfying $[\Sigma_K^-] \cdot [\Sigma^J_+] = 2 \delta^J_K$, and $P_{\a\, J}$, $Q_\a^K$ are integers or half-integers. Then the gauge coupling associated with each stack is 
\be
\frac{2\pi}{g_\a^2} =  P_{\a\, K}  u^K \, ,
\label{gaugec}
\ee
and the St\"uckelberg mass for the $U(1)$'s involved in the GS mechanism reads \cite{Ghilencea:2002da}
\be
M_{\a\b}^2 = 4 G_{KL} Q^K_\a  Q^L_\b g_\a g_\b \,  M_{\rm P}^2 \, .
\label{masstu}
\ee
Finally, supersymmetry implies that there is a D-term potential for the chiral fields charged under the massive $U(1)$'s, which reads
\be
V_D = \oh \sum_\a g_\a^2 \left(\xi_\a +  \sum_i q^i_\a K_{i\bar{i}} \Phi_i \bar{\Phi}_{\bar{i}} \right)^2    \, , \qquad \text{with} \quad \pi \xi_\a =   Q^K_\a\ell_K M_{\rm P}^2\, .
\label{VD}
\ee
Here $\Phi_i$ represents the scalar component of the 4d chiral field at two D6-brane intersections, with kinetic terms $K_{i\bar{i}}$ and charge $q^i_\a$ under $U(1)_\a$, and $\xi_\a$ is the Fayet-Iliopoulos term \cite{Cremades:2002te}, which is a linear combination of the dual saxions defined as $\ell_K = - \p_{u^K} K$. Interestingly, it was found in \cite{Casas:2024ttx} that the vev of such dual saxions may provide a rough estimate of the  gonion tower scale, via 
\be
m_{{\rm gon}, \a\b} \sim g_{\rm min} \ell_{\rm min}^{1/2} M_{\rm P}\, , 
\label{estimate}
\ee
where $g_{\rm min} = {\rm min} (g_\a, g_\b)$ and $\ell_{\rm min}$ is the smallest dual saxion involved in the FI-term of such $U(1)$. See \cite{Casas:2024ttx} for a more precise discussion and applications to specific setups.

	 \subsection{Yukawas at infinite distance}
	 
\begin{figure}[tb]
	\begin{center}
 \hspace{-3em}
	\includegraphics[scale=0.45]{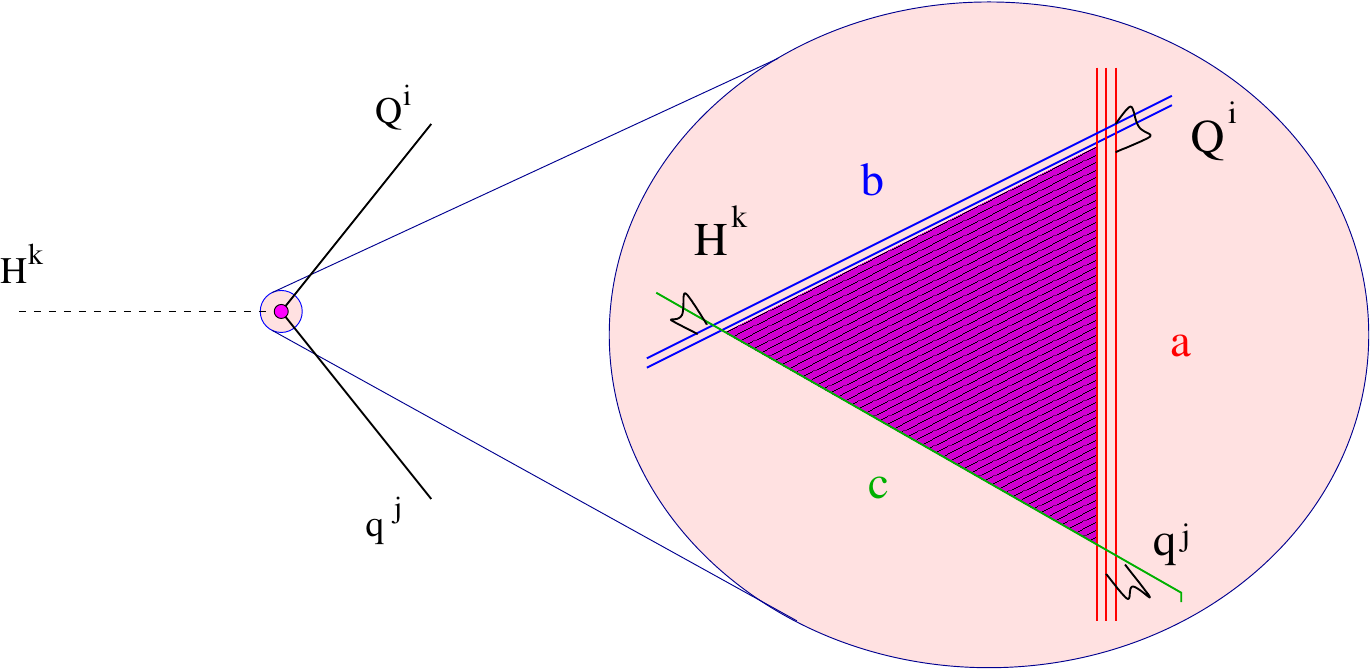}
	\caption{ Yukawa couplings arise at the intersection of three D6-branes.}
	\label{yukawa}
	\end{center}
	\end{figure}

	 Standard trilinear Yukawa couplings in type IIA CY orientifolds correspond to string amplitudes involving three D6-branes intersecting at angles at a single point in the CY, or connected through a worldsheet along a triangle joining three intersections of the involved D6-branes \cite{Aldazabal:2000cn}, see fig.(\ref{yukawa}). Those Yukawa couplings have been computed explicitly for toroidal or orbifold compactifications, see \cite{Cremades:2003qj,Cvetic:2003ch,Lust:2004cx,Bertolini:2005qh,Cremades:2004wa,DiVecchia:2008tm}.
	 The generic CY case is obviously more involved, but due to the locality of the interactions,
	 some general results for the Yukawa couplings can nevertheless be obtained.
	 Here we summarise some of the results in \cite{Casas:2024ttx} concerning Yukawa couplings at infinite field distance, in particular concerning the large complex structure limit in type IIA CY orientifolds. In this reference it was argued that canonically normalised Yukawa couplings among matter fields at D6-brane intersections,  labelled by the family indices $i,j,k$ are given by an expression of the form
	 \be
	 Y_{ijk}\, =\,  e^{\phi_4/2} {\rm Vol}_X^{1/4} W_{ijk} \Theta_{ijk}^{1/4}\, .
  \label{Ygeneral}
	 \ee
	 Here $W_{ijk}$ is the holomorphic superpotential coupling, which is a function of the K\"ahler moduli, and $\phi_4$ is the 4d dilaton. The functions $\Theta_{ijk}$ contain all the information about the local angles of intersection of the
	 different D6-branes, see e.g. ref.\cite{Casas:2024ttx}  for details and toroidal examples.  Finally ${\rm Vol}_X$ is the volume
	 of the compact CY three-fold $X_6$, which is a function of the K\"ahler moduli. 

  The challenge that we address in this work is to identify moduli space directions in which one can suppress the Yukawa couplings of neutrinos while maintaining unsuppressed (or not exceedingly suppressed) the rest of the SM Yukawas. In this sense, taking the limit $e^{\phi_4}{\rm Vol}_X^{1/2} = g_s \to 0$ is not a good strategy, since this factor affects all the Yukawas in a universal way.  A different procedure to get almost vanishing Yukawa couplings for the neutrinos is to construct models in which, by selecting certain points in K\"ahler moduli space the holomorphic Yukawas $W_{ijk}$ have some zero entries, which then get corrected by exponentially suppressed contributions from Euclidean E2-brane  charged instantons, see refs.\cite{Abel:2006yk,Blumenhagen:2007zk,Cvetic:2008hi,Ibanez:2008my,Blumenhagen:2009qh,Marchesano:2009rz,Anastasopoulos:2010hu}.
	 However, in practice it is quite challenging to find explicit D6-brane models in which {\it all} neutrino Yukawas become very small while avoiding that some other quarks or charged lepton Yukawas also become small, see some attempts in \cite{Cvetic:2008hi,Ibanez:2008my,Anastasopoulos:2010hu} and references therein.
  
	 In the following we will apply a different line of action, based on the approach and results of our recent work \cite{Casas:2024ttx}, that explored the structure of Yukawas at fixed CY volume and 
	 and large vevs for the complex structure fields. 
	 Geometrically, this corresponds to considering cases in which some of the intersection angles involved in a certain Yukawa couplings become very small \cite{Casas:2024ttx}.
   In those limits the structure of the said Yukawas reduces to the form
	 \beq 
	 Y_{ijk} \, \simeq \, B\,  e^{\phi_4/2} \Theta_{ijk}^{1/4}\, ,
  \label{YTheta}
	 \eeq
	 with $B$ taken to be an order one constant.
	 In the case that all three vertices have three gonion towers one can write \cite{Casas:2024ttx}
	 \beq
	 Y_{ijk} \, \sim \,  e^{-\phi_4} \left(\frac {m_{\rm gon}^i}{M_{\rm P}}\cdot\frac {m_{\rm gon}^j}{M_{\rm P}}\cdot\frac {m_{\rm gon}^k}{M_{\rm P}}\right)^{1/2} ,
  \label{Ymgon3} 
	 \eeq	
	 where the gonion scales correspond to the leading tower (if there is more than one) at each intersection. Along limits in which we send one or several saxions $u^K$ to infinity, we have that $e^{\phi_4} = M_s/M_{\rm P} \to 0$. Then, because we always have the hierarchy $m_{\rm gon} < M_s < M_{\rm P}$, \eqref{Ymgon3} tends to zero as well. These are the infinite distance limits that were explored in \cite{Casas:2024ttx} where it was argued that whenever $Y \to 0$ at least one of the gauge couplings involved in the Yukawa also tends to zero.
  
  There is a particularly interesting case in which only one intersection  (say, the $i^{\rm th}$)  presents small angles. In that case the expression is simplified to
	 \beq
	 Y_{ijk} \, \sim \,  e^{-\phi_4} \left(\frac{m_{\rm gon}^i}{M_{\rm P}}\right)^{1/2} \frac{M_s}{M_{\rm P}}
	 \simeq \   \left(\frac {m_{\rm gon}^i}{M_{\rm P}}\right)^{1/2} \ ,
  \label{Ygon2d}
	 \eeq	
	where we have used $M_s=e^{\phi_4}M_{\rm P}$. Thus in this case the Yukawa coupling is determined by the gonion scale of the	intersection with the smallest angle. This will be the case of interest to obtain realistic Dirac neutrino masses, as it will apply to the neutrino sector Yukawas. For the rest of the Yukawas, the idea is that they involve at least one ${\cal N}=2$ sector, see footnote \ref{ft:N=2}. Then, from \eqref{YTheta} one obtains $Y_{ijk} \sim h_{i\bar{i}}^{-1/2}   \left(\frac {m_{\rm gon}^i}{M_{s}}\cdot\frac {m_{\rm gon}^j}{M_{s}}\right)^{1/2}$ instead of \eqref{Ymgon3}, where $h_{i\bar{i}}$ is a function of the saxions $\{u^K\}$ bounded by $g_\a^{-1} g_\b^{-1}$, namely the two gauge couplings involved in the ${\cal N}=2$ sector \cite{Casas:2024ttx}. Because in this case there is no 4d dilaton suppression when expressing $Y_{ijk}$ in terms of ${m_{\rm gon}}/{M_{s}}$, these other Yukawas need not be suppressed along the infinite distance limits under consideration, allowing for a realistic set of couplings.

	 \subsubsection{A toroidal orientifold toy model}\label{toymodel}
	 
	To get a more detailed flavour of the structure of this limit we briefly consider here the toroidal orientifold example 
	described in \cite[section 4]{Casas:2024ttx}, see that reference for details and notation. It is obtained starting from a compact toroidal space
	$X_6=({\bf T}^2)_1\times ({\bf T}^2)_2 \times ({\bf T}^2)_3/\Gamma $, where $\Gamma = {\bf Z}_2\times {\bf Z}_2$ has the same action as considered in \cite{Cvetic:2001tj,Cvetic:2001nr}.  Each torus is parametrised by a complex coordinate $z_i=y_{2i-1}+i\tau_i \, y_{2i}$, $i=1,2,3$, where the geometric complex structure
	is $\tau_i=R_{2i}/R_{2i-1}$.  The untwisted complex structure moduli have real parts given by
	\beq
	s= \frac {e^{-\phi}}{4}R_1R_3R_5 \, , \quad   u^{(1)}  = \frac {e^{-\phi}}{4}R_1R_4R_6 \, , \quad
	 u^{(2)}  =\frac {e^{-\phi}}{4}R_2R_3R_6 \, , \quad  u^{(3)} =\frac {e^{-\phi}}{4}R_2R_4R_5,
	 \eeq 
	 with the 4d dilaton given by $e^{-2\phi_4} =4\sqrt{su^{(1)}u^{(2)}u^{(3)}}$, and the string scale reads $M_s=e^{\phi_4}M_{\rm P}$.
	 For $\tau_i\ll 1$  there are KK and winding towers 
	 determined by the scales respectively
	 \beq
	 m_{{\rm KK},i}\, =\, \frac {2M_s}{R_{2i-1}}\, =\, \frac {M_{\rm P}}{2A_i^{1/2}\sqrt{su^{(i)}}} \, , \qquad 
	 m_{{\rm w},i}\, =\, \frac {1}{2}R_{2i}M_s\, =\, \frac {A_i^{1/2}M_{\rm P}}{2\sqrt{su^{(i)}}}  \,  ,
	 \label{masillas}
	 \eeq
	 where $A_i=(R_{2i}R_{2i-1})/4$ is the area of $({\bf T}^2)_i$. We consider now stacks of $D6$-branes $D6_a$, $D6_R$, $D6_L$ 
	 wrapping the 3-cycles
	 \beqa
	 \Pi_a &=& 8(k,1)(k,1)(k,-1)\, , \\
	 \Pi_R &=& 2(0,1)(1,0)(0,-1)\, ,\\
	 \Pi_L &=&  2(0,1)(0,-1)(1,0) \, ,
	 \eeqa
	 with $k$ an integer. The $R,L$ branes are invariant under the orientifold and orbifold symmetries which leads to symplectic groups,
	 so that locally the gauge group has a Pati-Salam structure $U(4)\times SU(2)_R\times SU(2)_L$, with $k^2$ generations 
	 of `quarks an leptons' in the representations $F_R=({\bar 4},2_R,1)$,  $F_L=( 4, 1, 2_L)$, along with one Higgs field in the $H=(1,2_R,2_L)$ representation.
	 From the wrapping numbers, one can deduce the  gauge couplings which are given by
	 \beq 
	 \frac {2\pi}{g_a^2} \, =\, 2(k^3s\ +ku^{(1)} + ku^{(2)} + ku^{(3)})\, ,\qquad 
	 \frac {2\pi}{g_R^2} \, =\, 2u^{(2)} \, , \qquad  \frac {2\pi}{g_L^2} \, =\, 2u^{(3)} \, .
	 \eeq
	 There is a FI-term associated to the unique $U(1)_a$ boson
	 \beq
	 \xi_a \, =\,  \frac {M_{\rm P}^2Q_a^K}{4\pi}\frac{1}{u^{(K)}}\, =  \, -\frac {M_{\rm P}^2}{4\pi} \left(\frac {1}{s}  + \frac {k^2}{u^{(1)}} + \frac {k^2}{u^{(2)}} - \frac {k^2}{u^{(3)}}\right),
  \label{FItoy}
	 \eeq
  where we have used that $Q_a^0 = -1/2$ and $Q_a^1=Q_a^2=-Q_a^3 = -k^2/2$.
  
	 SUSY configurations correspond to $\xi_a=0$.  Let us now consider the limit $s,u^{(2)} = u$, $u\rightarrow \infty$ keeping areas fixed and $u^{(1)},u^{(3)}$ also bounded.
	  In that limit two dimensions decompactify along the second complex plane.  Since 
	 $su^{(i)} = e^{-2\phi_4}\tau_i^{-1}$ that limit corresponds to  $\tau_2\rightarrow 0$.  Then towers of gonions appear in the 
	 second complex plane for the  $aR$ and $aa^*$ intersections. The gonion masses in both cases are given by
	 \beq
	 m_{{\rm gon},2}  \simeq  \frac {M_{\rm P}}{2\sqrt{ksu^{(2)}}} 	 \simeq  \frac {M_{\rm P}}{k^{1/2} u} \, .
	 \eeq
	 The string scale in this limit goes as $M_s=e^{\phi_4}M_{\rm P}\simeq M_{\rm P}/u^{1/2}$ and the gauge couplings as $g_a\simeq g_R\sim 1/u^{1/2}$.  The gauge group $U(1)_a$ is anomalous, with a St\"uckelberg mass
	 \beq
	 M_V^2 = \frac {g_a^2M_{\rm P}^2}{2}\left( \frac {1}{s^2} 
	  + \frac {k^4}{(u^{(1)})^2} +  \frac {k^4}{(u^{(2)})^2}\ + \frac {k^4}{(u^{(3)})^2}\right) \, .
   \label{MVtoy}
	 \eeq
	 Note that in the limit $s,u^{(2)} \rightarrow \infty $ the vector mass is of order the string scale,
	 since $M_V\sim g_aM_{\rm P}\sim M_{\rm P}/u^{1/2}\sim M_s$, so there is no suppression of the St\"uckelberg mass.
	 
\begin{figure}[tb]
	\begin{center}\hspace{2em}
	\includegraphics[scale=0.275]{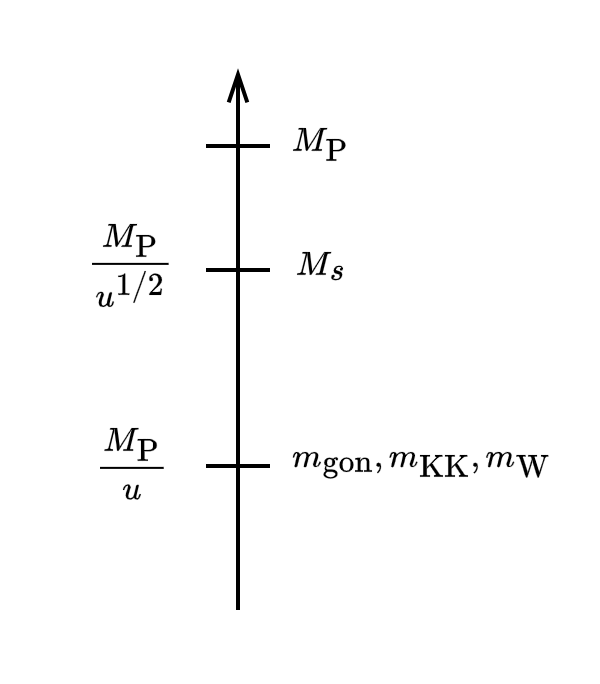}
	\caption{Structure of scales in the toy example in the text.}
	\label{escalillas}
	\end{center}
	\end{figure}

	 There is only one allowed Yukawa coupling  $F_R\times F_L\times H$. As we described above the magnitude of the Yukawa coupling is determined by the tower of gonions in the intersection where the right-handed fermions $F_R=({\bar 4},2_R,1)$ 
	 live, so that
	 \beq 
	 Y_{{F_R}F_LH} \, \simeq  \left(\frac {m_{{\rm gon},2}}{M_{\rm P}}\right)^{1/2} \simeq  \left(\frac {1}{\sqrt{ksu^{(2)}}}\right)^{1/2} \simeq  \frac {1}{u^{1/2}} \, .
	 \eeq
          So the relevant Yukawa coupling is suppressed in this limit as the gauge couplings of the fields $F_R$:
          \beq
           Y_{{F_R}F_LH} \, \simeq  \, g_a \simeq g_R \, .
           \eeq
          In  addition to the gonion towers along the 2$^{\rm nd}$ complex plane, there are KK and winding states corresponding to this plane and with masses given by eqs.(\ref{masillas}) so that in this limit
          \beq
          m_{{\rm KK},2} \, \simeq m_{{\rm w},2} \, \simeq \, \frac {M_{\rm P}}{u}  \, ,
          \eeq
          which is of the order of the gonion tower scale. Furthermore, there are also KK and winding states corresponding to the D6$_a$ and D6$_R$ branes.
          Thus for  D6$_a$ one gets KK and winding states of the form \cite{Casas:2024ttx}
          \beq
         M_{{\rm KK},\, {\rm D6}_{a,2}}\, \simeq \, \frac {M_{\rm P}}{A_2^{1/2}2k\sqrt{su^{(2)}} }\, \sim \, \frac {1}{u} \, , \qquad 
         M_{{\rm w,\, D6}_{a,2}}\, \simeq \, \frac {A_2^{1/2}M_{\rm P}}{2k\sqrt{su^{(2)}} }\, \sim \, \frac {1}{u} \, ,
         \eeq
          and similarly for the D6-brane $R$. All in all the structure of mass scales is shown in figure \ref{escalillas}.
          At a scale $\sim M_{\rm P}/u$ there is {\it i)} a tower of gonions transforming like $F_R$,  {\it ii)} towers of KK and winding states corresponding to two decompactified dimensions and  {\it iii)} towers of KK and winding states corresponding to open strings ending on the branes D6$_a$ and D6$_R$. The string scale is at an intermediate scale $M_s\simeq M_{\rm P}/u^{1/2}$.
          
          In this Pati-Salam setting unfortunately not only the neutrinos but all quarks and leptons get suppressed Yukawa couplings, which is not
          what we want. The idea is to consider D6-brane configurations in which one can isolate an intersection where 
          only the right-handed neutrinos live so that only the neutrino Yukawa couplings are suppressed.  We will discuss these types of configurations in the next section. Still the suppression mechanism, the structure of scales and the fact that Yukawa couplings scale with suppressed gauge couplings are  well exemplified by this toy model.

     \subsubsection{ Difficulties for a single large dimension or more than two}

   In the above  $s,u^{(2)}\rightarrow \infty $ limit two dimensions become large.  An interesting question is   how things change if only one dimension is large.  One can see that this is obtained in the limit
   $s,u^{(2)},t^{(2)}\sim u $,  $u\rightarrow \infty $, as explained in appendix \ref{onedimension}. Then
   the winding states remain at the string scale and only the KK tower becomes light.
As we describe in that appendix, getting a Yukawa coupling for neutrinos of the required size, implies 
the extra dimension must be extremely large, with $m_{KK}\simeq  10^{-57}$ GeV and a nonsensical Species Scale 
of order 100 eV. So making a single dimension large is not the way to get tiny
Dirac neutrino masses of the correct size.

A single large extra dimension has also received renewed interest within the context of the idea of `Dark Dimension' 
\cite{Montero:2022prj,Gonzalo:2022jac,Obied:2023clp,Anchordoqui:2023oqm}. In this case one has $m_{\rm KK}\simeq 10^{-3}$ eV, in possible connection with the scale
of the cosmological constant. As we said, as in all one large dimension models, one cannot obtain the required 
Dirac neutrino masses from tiny Yukawas. On the other hand we also show in Appendix \ref{onedimension}
that such a large dimension is 
compatible with the correct overall size for the SM Yukawa couplings only if along the 
large dimension there are no light SM gonion towers.

    Finally, one may consider limits in which four or six dimensions become large,
    see Appendix \ref{onedimension}. When applied to scenarios with small Dirac neutrino masses like the one in section \ref{s: SM intersecting}, it  gives rise to a too low quantum gravity cut-off that is ruled out experimentally.  Given these difficulties, in the rest of the paper we will focus on setups with two large geometric dimensions, as in the toy model above.

\section{ The SM at intersecting D6-branes }
\label{s: SM intersecting}
	 
Let us start by considering the simplest structure of D6-brane intersections which can reproduce the SM spectrum, see e.g. \cite{Blumenhagen:2005mu,Blumenhagen:2006ci,Marchesano:2007de,Lust:2009kp,Ibanez:2012zz,Marchesano:2022qbx,Marchesano:2024gul} and references therein.
	
The minimal set of $D6$-branes would involve  four stacks $a,b,c,d$ (and their orientifold images $a^*,b^*,c^*,d^*$), with multiplicities $N_a=3,N_b=2,N_c=N_d=1$, so that the naive gauge group reads $U(3)_a\times U(2)_b\times U(1)_c\times U(1)_d$. Instead of $N_b=2$ one can also consider $N_b=1$ if both $D6_b$ and $D6_b^*$ coincide and give rise to the gauge group $USp(2)\simeq SU(2)$. Let us first look at this simpler class of models. The gauge group will be $SU(3)\times SU(2)\times U(1)_a\times U(1)_c\times U(1)_d$. Here $U(1)_a$ and $U(1)_d$ correspond to gauged baryon number and (minus) lepton number respectively, while $U(1)_c$ corresponds to the diagonal component of right-handed weak isospin. The hypercharge is a linear combination of the three $U(1)$'s, while the other two become massive,
due to $B\wedge F$ couplings. At the intersection of D6-branes, i.e. at $X_4$ and a points in the compact space $X_6$, massless chiral multiplets appear. In particular, to have correct SM quantum numbers, the intersection numbers are given by \cite{Cremades:2002qm,Cremades:2003qj} 
\begin{equation}
\begin{split}
I_{ab} = I_{ab^*}  =  3\, ,   & \qquad     I_{ac} = I_{ac^*}  =  -3\, ,  \\
I_{db} = I_{db^*}  =  -3\, , & \qquad    I_{dc} = I_{dc^*}  =  3   \, ,   
\end{split}
\label{internumbers}
\end{equation}
with the rest of the intersection numbers vanishing. The hypercharge is given by the linear combination
\beq
 Y \,  = \,  \frac {1}{6}Q_a - \frac {1}{2}Q_c  +  \frac {1}{2} Q_d \, ,
 \eeq
whose gauge boson should remain massless (no $B\wedge F_Y$ couplings) in the spectrum.
The reader may check that each of the SM fields at the intersections have the correct 
hypercharge assignment.
There are in addition two other $U(1)$'s which in general will be massive.
Note that the right-handed neutrinos are localised at the intersection of the branes $c$ and $d^*$,
and are neutral under hypercharge, as they should.

As it stands this simple configuration is not appropriate for our purposes for two reasons. Indeed,  as we described in the previous section and in \cite{Casas:2024ttx}, in order to get tiny Yukawa couplings for the neutrinos, at least one of the gauge couplings $g_c$ or $g_{d}$ must be very small, corresponding to either the D6-branes $c$ or $d$ wrapping a three-cycle with large volume. However, then the hypercharge coupling $g_Y$ will also go to zero, which is incompatible with its experimental value $g_Y\simeq {\cal O}(0.3)$. Also, the charged leptons live at the $cd$ intersection and will also get a tiny Yukawa coupling, against the experimental results. Thus the desired brane configuration should differentiate between charged leptons and the 
right-handed neutrino sector.

A simple generalisation of this D-brane configuration that does the job can be obtained by adding an extra brane ${\tilde c}$ and also not imposing the condition $b=b^*$ on the electro-weak brane $b$. In fact, for simplicity, we will consider a D-brane configuration that is fully separated from its orientifold images and the orientifold planes, in the sense that all intersection numbers of the form $I_{ab*}$ and $I_{a{\rm O6}}$ vanish. In mirror type IIB setups, this feature is easily implemented by considering D3-branes at singularities away from the orientifold planes possibly with flavour D7-branes  \cite{Aldazabal:2000sa}. In the type IIA framework one should be able to engineer them via embedding the local models of \cite{Uranga:2002pg} into global CY orientifold compactifications. We do not claim that this is the unique way to build SM brane configurations with small Dirac neutrino masses, but this mirror picture provides us with an explicit framework, sufficient for the purpose of showing the minimal ingredients required to obtain the result. In appendix \ref{su5}, an $SU(5)$ configuration is also displayed in which the same structure to obtain small neutrino Yukawa couplings is shown.

\begin{figure}[tb]
	\begin{center}
 \hspace{1em}
	\includegraphics[scale=0.15]{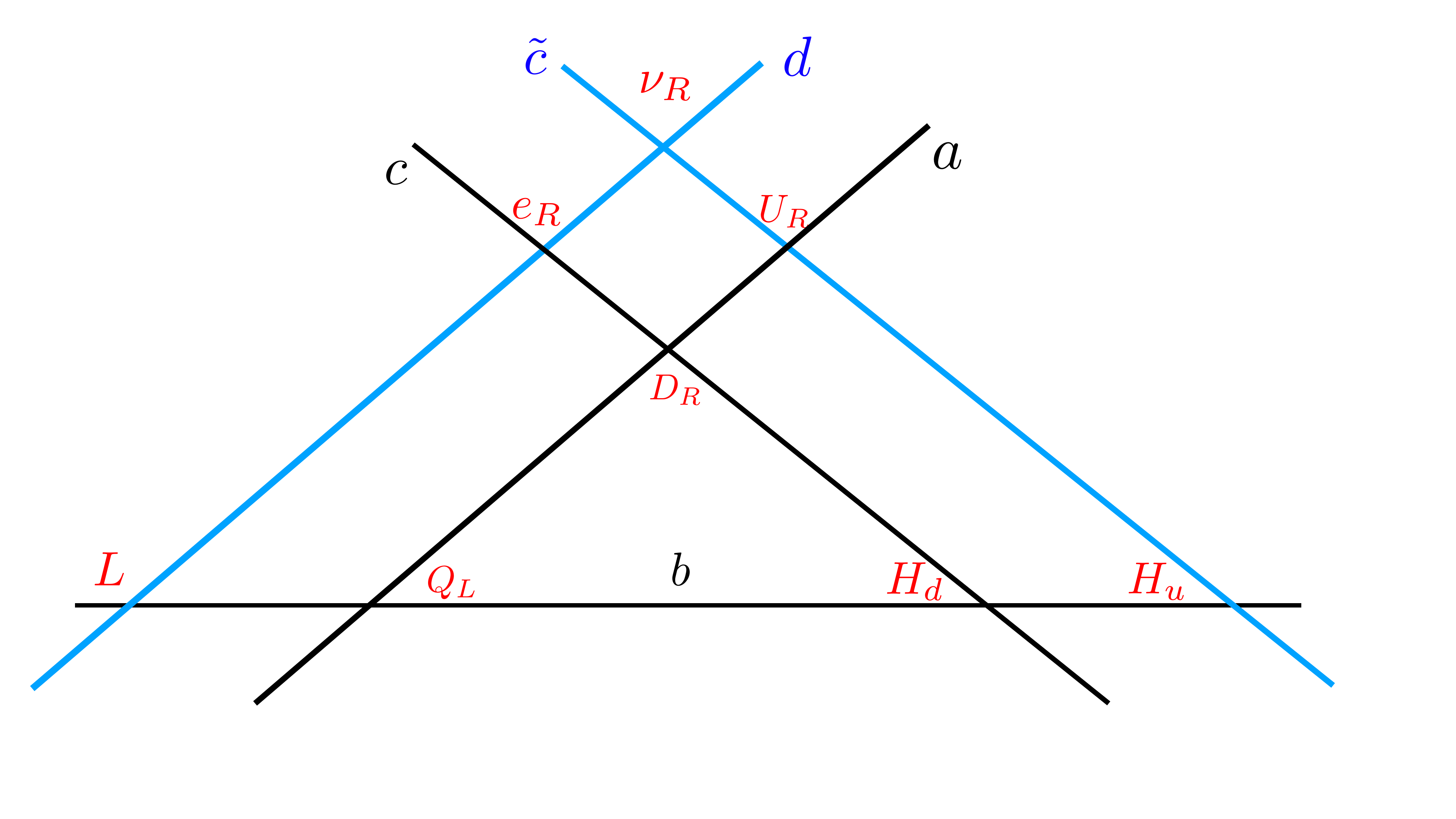}
	\caption{Scheme of the intersections in the SM-like quiver of table \ref{tablaSM-2}. The branes ${\tilde c},d$ in blue wrap a large volume, leading to suppressed Yukawa couplings for the right-handed neutrinos.}
	\label{smquiver}
	\end{center}
	\end{figure}

So let us consider a brane configuration with stacks $a,b,c,{\tilde c},d$ and intersection pattern described as in table \ref{tablaSM-2}. 
\begin{table}[h!!]\begin{center}
\renewcommand{\arraystretch}{1.00}
\begin{tabular}{|c|c|c|c|c|c|c|c|c|c|}
\hline
Intersection & Matter fields  & Charge  & $Q_a$  & $Q_b$  & $Q_c$  & $Q_{\tilde c}$ &  $Q_d$  & $Y$ & $Q_\nu$  \\
\hline\hline
$(ab)$ &  $Q_L$ &   $(3,2)$  &  1  & -1 & 0 & $0$ &  $0$  & 1/6 & 0\\
\hline
$(a{\tilde c})$ &  $U_R$ &   $({\bar 3},1)$  &  -1  & 0 & 0 & $1$ &  $0$  & -2/3 & 1 \\
\hline
$(ac)$ &  $D_R$ &   $({\bar 3},1)$  &  -1  & 0 & 1 & $0$ &  $0$  & 1/3 & 0\\
\hline
$(bd)$ &  $L$ &   $(1,2)$  &  0  & -1 & 0& $0$ &  $1$  & -1/2 & -1 \\
\hline
$(cd)$ &  $E_R$ &   $(1,1)$  &  0  & 0 & 1 & $0$ &  $-1$  & 1 & 1 \\
\hline
$({\tilde c}d)$ &  $\nu_R$ &   $(1,1)$  &  0  & 0 & 0 & $1$ &  $-1$  & 0 & 2 \\
\hline
$(b{\tilde c})$ &  $H_u$ &   $(1,1)$  &  0  & 1& 0 & $-1$ &  $0$  & 1/2 & -1\\
\hline
$(bc)$ &  $H_d$ &   $(1,1)$  &  0  & 1& -1 & $0$ &  $0$  & -1/2 & 0 \\
\hline			
\end{tabular}
\caption{SM spectrum and  D-brane $U(1)$ charges in the configuration with an five branes \eqref{gaugeG}. The intersection numbers for the Quark-Lepton sector are $I_{ab} = - I_{ac} = - I_{a\tilde{c}} =  - I_{bd} = I_{cd} = I_{\tilde{c}d} = 3$, and for the Higgs sector $I_{bc} = I_{b\tilde{c}} = 6$. }
\label{tablaSM-2}
\end{center}
\end{table}
D-brane configurations with similar structure have been considered previously in e.g.\cite{Aldazabal:2000cn,Verlinde:2005jr,Antoniadis:2021mqz}. The gauge group is of the form
\beq 
SU(3)\times SU(2)\times U(1)_a\times U(1)_b\times U(1)_c\times U(1)_{\tilde c}\times U(1)_d \, .
\label{gaugeG}
\eeq
Most $U(1)$'s here are anomalous and will get a mass via $B \wedge F$ couplings implementing a St\"uckelberg mechanism. The hypercharge is now identified with
\beq
Y \, =\, \frac {2}{3}Q_a + \frac {1}{2} Q_b  + Q_c\, ,
\label{hyperQ}
\eeq
and it should be free of St\"uckelberg couplings to axions. In appendix \ref{anomalias} we describe the $U(1)$ anomaly structure of this setting and the general conditions that should be obeyed 
in a specific CY orientifold so that all 
$U(1)$'s but hypercharge become massive and the $U(1)_{\tilde c}$, $U(1)_d$ couplings may become small. Figure \ref{smquiver}  symbolically summarises the intersections which are assumed. In the figure, any closed triangle corresponds to a Yukawa coupling allowed by the $U(1)$ charges.  

Note that the simpler configuration in \eqref{internumbers} is recovered if we identify the branes  ${\tilde c}$ with $c^*$  and replace the $U(2)$ of brane $b$ by an $USp(2)$. The essential new ingredient that the present configuration brings is that now the right-handed neutrinos are localised  at the intersection of branes ${\tilde c}$ and $d$ whose associated $U(1)$'s do not appear in the expression for the hypercharge. Thus their associated gauge couplings $g_{\tilde c},g_d$ may tend to zero without affecting any of the SM gauge couplings.

Let us use the type IIB mirror picture of D-branes at singularities to illustrate how all the features described above can be achieved in a single construction. Let us first assume that the D-branes $a$, $b$, $c$ correspond to D3-branes at a singularity away from the orientifold, that reproduces the chiral content of this subsector. From here it follows that \eqref{hyperQ} is free of any $B \wedge F$ coupling and hence massless. The branes $\tilde{c}$ and $d$ are instead assumed to be flavour D7-branes. This will generically imply that all additional anomaly-free $U(1)$'s that arise from \eqref{gaugeG} develop $B \wedge F$ couplings. Moreover, it will imply that all the SM Yukawas involve an ${\cal N} = 2$ sector and are insensitive to the bulk moduli, except those that involve right-handed neutrinos. Indeed, the $\nu_R$ arise at the intersection of two flavour D7-branes and far away from the singularity. As such, their gonion spectrum will be sensitive to the CY bulk moduli, and so will be their Yukawas. 

\subsection{The neutrino Yukawa coupling}

Let us now consider the Yukawa coupling for the neutrinos:
\beq
{\cal L}_\nu \, =\, Y_{\nu, ij}\, H_u L^i\nu_R^j \, .
\label{YuknuR}
\eeq 
Dirac neutrino masses arise when the Higgs field $H_u$ acquires a vev, and we want to study the conditions under which these masses become small, of order $(10^{-1}-10^{-4})$ eV. As we described in the last section, in limits of large complex structure a small Yukawa is accompanied with a light gonion tower. In our setup, one expects that only the sector $\tilde{c}d$ has light gonions, since it corresponds to the intersection of the two `flavour' branes. Moreover, it was argued in  \cite{Casas:2024ttx} that when a gonion tower at the intersection $\tilde{c}d$ asymptotes to zero mass in Planck units, at least one of the gauge couplings $g_{\tilde{c}}, g_d$ also goes to zero. In other words, we have that the imaginary part of the complex structure field
\be
U \equiv \left(P_{\tilde{c},K}+P_{d,K}\right) U^K \, ,
\label{defU}
\ee
tends to infinity, since
\be
u \equiv  \im U = \frac{2\pi}{g_{\tilde{c}}^2} +  \frac{2\pi}{g_{d}^2}\, .
\label{defu}
\ee
The limits of interest will thus be those in which $u \to \infty$ and the rest of the gauge couplings $g_a$, $g_b$ and $g_c$ remain constant, so that the hypercharge coupling is unaffected along the limit. 

The complex structure saxion \eqref{defu} can be interpreted as the gauge coupling of the  gauge group $U(1)_{\nu}$ felt by the right-handed neutrinos, which corresponds to the combination coupling to the generator 
\beq
Q_\nu \, \equiv \, Q_{\tilde c}  - Q_d  \, ,
\label{Qnu}
\eeq
so that $Q_\nu(\nu_R^i)= 2$. This gauge group will be anomalous (see appendix \ref{anomalias}) and hence should be massive due to a $B\wedge F$ coupling of the form
\be
S_{BF_\nu}\, =\, Q_\nu^K \, B_K\wedge F_{\nu} \, \equiv\, \left(Q_{\tilde{c}}^K-Q_d^K\right)\, B_K\wedge F_{\nu}  \, ,
\label{BFnu}
\ee
where the $2Q_{\tilde{c},d}^K$ are integers determined by the  wrapping numbers of the D6-branes $\tilde{c}$ and $d$, respectively, that represent the above $U(1)$ generators, see \cite{Casas:2024ttx} and the discussion of section \ref{s:orientifolds} for their precise definition. The coupling \eqref{BFnu} determines the way in which the chiral multiplet associated to $U$ and the vector multiplet $V_\nu$ that corresponds to \eqref{Qnu} combine in the K\"ahler potential $K$. In particular, the Green--Schwarz mechanism fixes that these two multiplets appear in the combination $M_{\rm P}^{-2} K =  k \left(u -  I_{\tilde{c}d} V_\nu, \dots \right)$, where $k$ is a real function of $u$ and other complex structure saxions. In the regime in which $u \to \infty$ we can approximate this function as 
\beq 
	M_{\rm P}^{-2} K \, \simeq \, -  \, n \log (u - I_{\tilde{c}d}  V_{\nu}  ) - \log (g(\{u^I\})) + \dots \, ,
	\label{largeM}
	\eeq
where $1 \leq n \leq 4$,  $g$ is a polynomial on a subset $\{u^I\}$ of the complex structure saxions $\{u^K\}$ and the dots represent subdominant terms of the K\"ahler potential in this regime. Along this limit the kinetic term for \eqref{Qnu} behaves as
\beq
\re f_\nu \, \simeq \, u  \implies   g_\nu \, \sim  \frac {1}{u^{1/2}} \to 0 \, .
\eeq
Additionally, associated to the  $B \wedge F$ coupling \eqref{BFnu} there is an FI-term that reads 
\be
\xi_\nu \simeq  Q_{\nu}^K  \ell_K  \, M_{\rm P}^2 \, . \label{eq: FInu}
\ee
If all the indices $K$ that appear in this expansion are contained in the subset of complex structure saxions $\{u, u^I\}$ that appear in the leading term of the K\"ahler potential \eqref{largeM},\footnote{In the language of \cite{Lanza:2021udy} this means that the limit is non-degenerate with respect the subset $\{u, u^I\}$.} then one can show that $\ell_{\rm min} \simeq 1/u$, where $\ell_{\rm min}$ is defined as in \eqref{estimate}. Then, applying such an estimate, one obtains that 
\beq
m_{{\rm gon},\nu} \, \sim \, \frac {M_{\rm P}}{u} \, \sim \, g_\nu^2M_{\rm P} \, .
\eeq

Let us consider the scaling of the Yukawa coupling in the large $u$ limit. As described in section \ref{s:orientifolds} in principle one gets the structure 
\beq
Y_{\nu,ij} \, \simeq \,  e^{-\phi_4} 
\left(\frac { m_{{\rm gon},\nu}^i }{M_{\rm P}}\right)^{1/2}\left(\frac { m_{{\rm gon},L}^j}{M_{\rm P}}\right)^{1/2}\left(\frac { m_{{\rm gon},H_u} }{M_{\rm P}}\right)^{1/2} \, .
\label{Yijnu}
\eeq
However, if we follow our mirror type IIB intuition in which the branes $a$, $b$, $c$ are represented by D3-branes at a singularity and $\tilde{c}$, $d$ are `flavour' D7-branes, we only expect the angles at the intersection $\tilde{c}d$ to be directly sensitive to the bulk modulus $u$, and so $m_{{\rm gon},\nu}$ to be the lightest tower. The other towers will be less relevant for determining the asymptotic behaviour of the Yukawa, although they may play a role in the details of the flavour dependence (along with the holomorphic K\"ahler dependent factor $W_{ijk}$).  For the heaviest neutrino generation (which we take as the 3$^{\rm rd}$ one) one expects $m_{{\rm gon},L }^{3{\rm rd}}\simeq m_{{\rm gon},H_u}\simeq M_s\simeq e^{\phi_4}M_{\rm P}$, so that one gets
\beq
Y_{\nu,3} \, \simeq \, 
\left(\frac { m_{{\rm gon},\nu}^i }{M_{\rm P}}\right)^{1/2}\, \simeq \, g_\nu \, .
\label{remarkably}
\eeq
This is a remarkably simple result, which allows us to fix all the relevant mass scales in the theory, as we describe below. Many of the features of our scenario reproduce those in the toy model of section \ref{toymodel}, because physical Yukawa couplings are, to leading order, only sensitive to the local geometry around D-brane intersections. Notice that the main ingredient of this construction is a field trajectory of infinite distance that leaves the hypercharge coupling invariant but sends $g_\nu \to 0$. Thus we believe that is a general result which should apply to most SM-like D-brane in type II CY orientifolds, whenever hierarchically small Dirac Yukawa couplings for neutrinos is pursued.

Note that this behaviour of the Yukawa coupling may be traced back to the fact that the field metric $K_\nu^i$ (in  units of $M_s$, see \cite{Casas:2024ttx} for details) for the right-handed neutrinos at the intersections becomes singular as the $\nu_R$ gonions become light, i.e.
\beq
K_\nu^i \, \simeq \, \frac {M_{\rm P}}{m_{{\rm gon},\nu}^i} \, \sim \, u \, .
\eeq
From the supergravity equation for canonically normalised Yukawas
\beq
Y_{ijk} \ =\ {W}_{ijk} \,\,e^{K/2}\left(K_iK_jK_k\right)^{-1/2} \ ,
\eeq
the behaviour $Y_\nu \sim K_\nu^{-1/2}\sim u^{-1/2}$ is inherited. 

While neutrino Yukawa couplings behave like $Y_{\nu}\sim u^{-1/2}$, and produce tiny Dirac neutrino masses, one may wonder whether large right-handed Majorana masses could also be present at the same time. They could in principle be generated by charged D-brane instantons, as mentioned in the introduction. This would be bad news, because the resulting neutrino masses would be too small. However, due to the constraints on the intersection numbers of such D-brane instantons with the SM branes \cite{Ibanez:2006da}, one expects that the instanton action depends on the field $U$. From where one obtains a Majorana mass suppression of order $e^{-u}$ which deactivates the see-saw mechanism. 

\subsubsection*{The species scale}

One of the results of \cite{Casas:2024ttx} is that when  some gonion towers become light, some dimensions of the compactification manifold become large. As already mentioned, the case with a single large extra dimension is in tension with light gonion towers, so let us consider that two internal dimensions become large along the limit $u \to \infty$ just like in the toy example of section \ref{toymodel}.  These two towers of KK/winding states have masses of the same order as the gonions: $m_{{\rm KK}}\sim m_{{\rm w}}\sim M_{\rm P}/u$.  In addition there can be KK-like towers coming from open strings at the worldvolume of the D6$_{\tilde c}$ and/or D6$_d$ branes, also with masses of this order. 

Along this complex structure limit the string scale also falls polynomially with $u$. One way to estimate its scale is to think that since we are probing infinite distance limits with open string states, one would expect the string scale to be the quantum gravity cut-off of the theory. In fact, this expectation agrees in toroidal models, where the quantum gravity cut-off is computed in different instances (see \cite[section 5]{Casas:2024ttx}). The quantum gravity cut-off, oftentimes dubbed the species scale \cite{Dvali:2007hz,Dvali:2007wp,Dvali:2008ec}, is the scale that emerges for a large number of species well below the Planck scale and where the effects of quantum gravity are not innocuous. It may be defined in several ways, and its value is of order 
\be
\Lambda_{\rm QG}\simeq \frac{M_{\rm P}^{D}}{N^{1/(D-2)}},\label{eq: lambda}
\ee
with $D$ the number of spacetime dimensions and $N$ the number of fields below $\Lambda_{\rm QG}$. For the case at hand, the number of species is given by $\Lambda_{\rm QG}^2\simeq \, N \,m_{{\rm light}}^2$,\footnote{Note the factor of 2 in the exponent, indicating that two multiplicative towers of states become light. In general, $p$ towers are approximated by an  effective single tower with $m_n=n^{1/p}m_{{\rm tower}}$ \cite{Castellano:2021mmx}. } and  after substituting into equation \eqref{eq: lambda} we arrive at 
\be
 \Lambda_{\rm QG}\simeq M_s \simeq \frac {M_{\rm P}}{u^{1/2}} \ \simeq \ g_\nu M_{\rm P} \,.
\ee

So all in all the structure of scales is again as in figure \ref{escalillas}. At the lowest scale with a mass $\sim g_\nu^2M_{\rm P}$ there is the tower of $\nu_R$-like gonions, KK/winding states of two extra dimensions, and the corresponding KK/winding states of the bulk of the large branes D6$_{\tilde c}$ and D6$_d$. Then the string scale is located at the geometric mean of that scale and the Planck scale: $M_s\simeq g_\nu M_{\rm P}$.

Finally, one can apply the same analysis to limits with four or six large dimensions, as performed in appendix \ref{onedimension}. There one finds that the relation becomes $\Lambda_{\rm QG} \sim g_\nu^{\frac{2p}{p+2}} M_{\rm P}$, where $p$ is the number of large dimensions. When combined with \eqref{remarkably} and the experimental data for neutrino masses, this gives an unacceptably low species scale.

\subsection{The light vector boson mass}

An important point to address is to determine the mass of the massive boson $V_{\nu}$ that couples to the right-handed neutrinos. The point is that
extra massive $U(1)$ bosons beyond hypercharge are generally given a mass that is essentially determined by the size of their FI terms.
So most of these $U(1)$ will have masses on the order of the string scale, if their FI terms do not depend on $u$.  However, in our setup such a dependence is expected, given the presence of gauge couplings in \eqref{masstu}, and in particular of $g_\nu \sim u^{-1/2}$. Because from the K\"ahler potential we will also have some kinetic terms that scale as $G_{uu} \sim u^{-2}$, in principle one could have a St\"uckelberg mass for $V_\nu$ that scales as $M_{V_\nu} \sim g_\nu^3 M_{\rm P}$. Such a gauge boson would then be the lightest massive particle in the spectrum.  With $g_\nu\simeq Y_\nu$, that would correspond to a vector boson with a mass $M_{V_\nu}\simeq  10^{-9}$eV and a coupling $g_\nu\simeq 10^{-12}$, which would be already ruled out by experiments measuring deviations of Newton's laws at short distances.

	In fact there are several contributions to the mass of this vector boson, which makes it typically much heavier. First, there is a model independent lower limit for its mass 
	\beq
	M_{V_\nu} \, \gtrsim \, m_{\nu,3}\,  \simeq \, 0.1\, {\rm eV} \, ,
	\eeq
	which comes from the fact that the Higgs boson $H_u$ is charged under $Q_\nu$. Once the Higgs gets a vev the vector boson gets a contribution $\delta(M_{V_\nu}(\text{Higgs}))\simeq g_\nu |\langle H_u\rangle|\simeq m_{\nu,3}$.
	Another possible, but {\it model dependent } contribution may arise if some right-handed sneutrino gets a vev. Then one would get
	$\delta(M_{V_{\nu}}(\text{sneutrino}))\simeq g_\nu |\langle\nu_R^i\rangle|\lesssim g_\nu M_s $. This would be of the order of the $\nu_R$-gonion  tower. 
	
	Finally, in general there are additional contributions to $M_{V_\nu}$ that come from \eqref{masstu}, besides the one that we have already accounted for. These come from considering that the vector $Q^K_{\nu}$ is not proportional to the field direction $U$ defined in \eqref{defU}. To be more concrete, let us in particular consider that $Q^K_{\nu}$ has entries pointing along some of the fields $u^I$ that appear in \eqref{largeM}, that do not grow along the limit. Then the total	St\"ukelberg mass for the gauge boson $V_{\nu}$ will be of the form
	\beq
	M_{V_\nu} \,  \simeq \,  g_\nu^2 M_{\rm P}^2 \left( \frac{I_{\tilde{c}d}^2}{u^2} +  4 G_{IJ} Q^I_\nu Q^J_\nu \right) \, ,
	\eeq
	where $G_{IJ} = - \oh \p_{u^I} \p_{u^J} \log g(\{ u^I\})$ is a homogeneous function of degree minus two the on the fields $u^I$, thus independent of $u$. In the limit $u\gg u^I$ we can take $G_{IJ} Q^I_\nu Q^J_\nu$ to be an order one numeric factor, and so the mass of the vector boson is of order $M_{V_\nu} \simeq g_\nu M_{\rm P}\simeq M_s$, namely of order the string scale.  Note that this is precisely what happens in the toy model of section \ref{toymodel}, where the vector $Q_a^K$ corresponding to the brane $a$ has non-vanishing entries along all fields, from where \eqref{MVtoy} follows. Of course one can still get a light vector boson by considering a more involved limit in which all the $u^I$ that couple to $Q^I_\nu$ go to infinity. However, then typically one obtains that other gauge couplings (like SM couplings) would also get very small.  
 
To sum up, the St\"uckelberg mass of the vector boson in $V_\nu$ is quite model dependent, in the sense that is quite sensitive to the integer entries $2Q_\nu^I$ of the D-brane model. In terms of the 4d EFT, what these integers specify is the field dependence of the FI-term associated to $V_\nu$. In general, one can only state that the
	$U(1)_\nu$ gauge boson may have a mass in the region
	\beq
	 m_{\nu,3} \, \lesssim \, M_{V_\nu} \, \lesssim \, g_\nu M_{\rm P}\, \simeq \, M_s \, ,
	 \eeq
	with the largest values being more generic.	In the low range of masses the vector boson mass is strongly constrained experimentally, as we briefly discuss below.

\section{Dirac neutrino masses: all scales fixed}
\label{phenoresults}

           From the previous discussions it is clear that in our setup all scales appearing in the theory are determined by a single parameter $g_\nu$. 
           Luckily, we can draw information about its value from experimental information on neutrino masses and mixings. 
           To make an estimate let us assume normal hierarchy of neutrino masses. Similar results would be obtained for an inverted hierarchy. 
           Let us recall from \eqref{Ygeneral} and  \eqref{YTheta} that the complete Yukawa couplings of neutrinos ${\tilde Y}_\nu$
           (including now for completeness the dependence on the K\"ahler moduli) will have the form
           \be
           {\tilde Y}_{\nu,i} \, \simeq \, Y_{\nu,i} \, \delta ^i\, ,\qquad   i=1,2,3 \, ,
           \ee
           where we have taken diagonal Yukawas for simplicity. Also, $\delta^i$ may be flavour dependent constants, analogous to those for charged lepton-fermion matrices, which are expected to give tiny corrections compared to the leading factor of $Y_{\nu,i}$ discussed in the last section. Let us consider two 
           limiting possibilities:

           {\it i) Universal $\nu_R$ gonion towers}
           
           In toroidal examples and
           in general compactifications without strong local curvature effects, one expects $Y_{\nu,i}=Y_\nu$,
           which corresponds to the three neutrino intersections having their corresponding gonion masses at the
           same scale (equal angles). In this case for the heaviest third generation neutrino one expects 
           $\delta\simeq 1$, so one has 
           \be
           {\tilde Y}_{\nu,3} \, \simeq \, Y_{\nu}\,  g_\nu\, ,  \qquad  i=1,2,3 \, ,
           \ee
           as already discussed in the previous section.
Moreover, for the third (heaviest) neutrino 
         generation one has
          \be
          m_{\nu,3} \, =\, Y_{\nu,3} \langle|{ H_u}|\rangle \, .
          	\ee
	  Then, the mass of the heaviest neutrino can be estimated from the neutrino oscillation data, i.e. $m_{\nu,3}\simeq \sqrt{\Delta m_{32}^2}\simeq 5\times 10^{-2}$ eV. From here one obtains
	 \be
	 Y_{\nu,3} \, \simeq \, g_\nu \, \simeq  7 \times 10^{-13} \, .
	 \ee
	Using this value for the gauge coupling $g_\nu$, one can  then calculate the different numerical scales that arise,
	which are shown in table \ref{tablaEscalas}.

		\begin{table}[h!!]\begin{center}
			\renewcommand{\arraystretch}{1.00}
			\begin{tabular}{|c|c|c|c|c|c|}
				\hline
			String Scale 	 &   SM gonions &   $\nu_R$ tower &   large dim  & Vector boson  & Gravitino  \\
				\hline\hline
				$M_s$ &  $m_{\rm gon}^{\rm SM}$ &   $m_{\rm gon}^\nu $  &  $m_{\rm{KK/w}}$  &  $M_{V_\nu}$ & $m_{3/2}$\\
							\hline
					$g_\nu M_{\rm P}$  &  $\lesssim M_s $ &   $g_{\nu}^2M_{\rm P}$ &  $g_\nu^2M_{\rm P}$ & $g_\nu |{\bar H}|-g_\nu M_{\rm P}$ &$\lesssim M_s^2/M_{\rm P}$ \\
					\hline
							$g_\nu=Y_{\nu,3}$\ , \ 700\ TeV & $\lesssim 700 $ TeV & $500$ eV & $500$ eV & $0.5 $ eV- $ 700\ $TeV  & $\lesssim \ 500$ eV \\
							\hline		
       $g_\nu=Y_{\nu,1}$\ , \ 10\ TeV & $\lesssim 10 $ TeV & $0.1$ eV & $0.1$ eV & $10^{-3} $ eV- $ 10\ $TeV  & $\lesssim \ 0.1$ eV \\
							\hline	
			\end{tabular}
			\caption{Spectrum of masses and scales from imposing Dirac character to neutrino masses in string theory.
   Numerical results are shown for two limiting cases with $g_\nu\simeq Y_{\nu,3} \simeq 7\times 10^{-13}$ and   $g_\nu \simeq Y_{\nu,1}\simeq 10^{-14}$.}
			\label{tablaEscalas}
		\end{center}
	\end{table}

  {\it ii) Non-universal $\nu_R$ gonion towers}

  Another possibility is that the local intersection angles of the three generations are not universal so that the  angle of the lightest generation $\nu_R$ is $\theta^{1} \ll \theta^{2,3}$. This may happen when the D6-brane intersection are around regions of $X_6$ with strong curvature, breaking universality.  As a limiting case, we may consider  the situation in which the lightest generation got its smallest Yukawa coupling from $Y_{\nu,1} $ (i.e. $\delta^1=1$).
  In that case one has
\be 
	 Y_{\nu,1} \, \simeq \, g_\nu \, .
	 \ee
The mass of the lightest neutrino is experimentally unknown. Cosmological bounds
suggest \cite{DiValentino:2021hoh,GAMBITCosmologyWorkgroup:2020rmf}
that $m_{\nu,1}\lesssim 20-40\times 10^{-3}$ eV and the bounds based on Swampland arguments mentioned below
yield $m_{\nu,1}\lesssim 7\times 10^{-3}$ eV
\cite{Ibanez:2017kvh,Gonzalo:2021zsp,Hamada:2017yji}. We show the numerical results in this case assuming for comparison $m_{\nu,1}\simeq 1\times 10^{-3}$ eV. We then have $g_\nu\simeq Y_{\nu,1}\simeq 10^{-14}$.
The corresponding structure scales is shown in table \ref{tablaEscalas}.

 Let us comment on the different masses at the different scales.

 \begin{itemize}

 \item {$M_s\simeq g_\nu M_{\rm P}\simeq 10-700$ TeV}.
 This is the string scale, which is the fundamental scale of the theory.

 \item{ $m_{\rm gon}^{\rm SM}\simeq m_{\rm KK}^{\rm D6}\simeq m_{\rm KK}\ \lesssim g_\nu M_{\rm P} \simeq 10-700 $TeV}.
  Below the string scale there are several types of 
 particles: 1) There are the gonions and vector-like copies of the SM fermions, which are expected to be below $M_s$. In fact, eventually 
 SUSY has to be broken, so squarks and sleptons may be considered as particularly light gonions. 
  2) There are KK  copies of the SM gauge bosons and their gaugini.  Again,  after SUSY breaking there will also be the
  gauginos of the SUSY SM.
  3) In general there will be additional massive $U(1)$'s, like those that are anomalous but not suppressed, with masses in this region. 
  4) The graviton KK copies corresponding to
 four (not so large) dimensions. 
 
\item { $m_{\rm gon}^{\nu_R}\simeq m_{\rm KK/w}^{\rm large}\simeq  m_{\rm KK/w}^{Q_\nu}\simeq  g_{\nu}^2M_{\rm P}\simeq 0.1-500$ eV}.
At this scale there are three types of particles again: 1) The gonion tower with copies of 
right-handed neutrinos. They come along with their sneutrinos if SUSY-breaking is not felt strongly 
by this sector; 2) The KK/w  copies of the gauge boson which couples to $Q_\nu$, and its gaugini;
3) The KK/w towers of the two large dimensions. 

\item
{$M_{V_\nu}\simeq  10^{-3}-0.1\ {\rm eV}$  \ to \ $10-700\ $TeV}.
The vector boson associated to the generator $Q_\nu$ may be the lightest new particle in the setting 
(along with its gaugino). As we said the vector gets a tiny model-independent St\"uckelberg mass 
of order $g_\nu^3$  eV.  However there is the  Higgs boson contribution to the mass
which is of order neutrino masses. Furthermore, if additional St\"ukelberg couplings to $Q_\nu$ exist,
the vector boson may be as massive as the string scale $\sim 10-700$ TeV.

\item $m_{3/2}\lesssim g_\nu^2M_{\rm P}$.
  Since the fundamental cut-off is at the string scale $M_s\simeq g_\nu M_{\rm P}$, the gravitino mass 
  is expected to have a mass $m_{3/2}\lesssim M_s^2/M_{\rm P} \simeq g_\nu^2 M_{\rm P}\simeq 0.1-500$ eV.
  Note that then supergravity mediated SUSY breaking mass terms will be tiny, of order $0.1-500$ eV, and hence
  they cannot be the leading source of SUSY breaking in the SM sector.

\end{itemize}

To this novel spectrum beyond the SM one has to add the modulus $U$ itself, whose vev determines all hierarchies. One expects that some dynamics fixes its value at $u \simeq g_\nu^{-2}$. In particular, it is conceivable that the above mentioned tiny soft terms
from supergravity mediation could induce a mass of order $0.1-500$ eV for this modulus.
We will not consider here the possible mechanisms that could give rise
to this modulus stabilisation, nor the stabilisation of the other four compact dimensions.

\subsection{Some phenomenological consequences and constraints}

 In this section we just give a preliminary overview of the possible observable consequences 
 of such a spectrum. An incomplete list includes:

 \begin{itemize}

\item {\it Constraints on the vector boson mass/coupling:} There are experimental limits in the plane 
gauge-coupling versus mass for `dark photon' $U(1)$'s beyond the standard model. They mostly arise from astrophysical constraints coming from over-cooling of stars, mainly the Sun, red giants, and horizontal branch stars, see e.g.\cite{Li:2023vpv,Heeck:2014zfa} for a recent discussion and references. 
There are also constraints from fifth-forces. Thus e.g. in those references constraints for a vector boson coupled to $(B-L)$ are presented. They find that e.g. a gauge coupling $g_{B-L}\simeq 10^{-12}-10^{-13}$ and a mass $M_{B-L}\simeq 1$ eV are barely
consistent with the bounds. It would be interesting to make a similar analysis with the gauge group 
which couples to our generator $Q_\nu$. The latter is not orthogonal to $U(1)_{B-L}$ so we would expect 
similar limits in the coupling/mass plane. Thus the presence of this model-independent `dark photon' in the theory may provide interesting experimental constraints.

\item {\it Constraints on KK and string scales:}
There are experimental constraints on the number and size of extra dimensions (see \cite{Antoniadis:2023doq} for a recent review and references).
 In our case 
we have two large dimensions with a scale $\sim 1/ (500\ {\rm eV})$, while the other four are of the order of the 
string scale $1/M_s$. Tests of the gravitational force at sub-millimeter scales give the bound
$M_s\geq 4.0$ TeV \cite{ParticleDataGroup:2022pth}. Searches of string resonances in ATLAS and CMS at LHC 
give a bound $M_{s}\gtrsim 8$ TeV. There are stronger astrophysical limits under the 
assumption that the KK states decay with a substantial branching ratio to photons, but that is not the case for the two large dimensions in question, which are isolated from SM couplings.  

These limits are below $M_s\simeq 10-700$ TeV which we have in the Dirac neutrino setting here described, but close to possible tests at LHC and future colliders. There are also constraints on KK SM gauge bosons, of order a few TeV.

\item {\it Neutrino masses}

The most obvious  prediction is the Dirac character of neutrinos. Another interesting aspect is the mixing of neutrinos with the neutralinos of the $U(1)_{\nu}$ gauge boson. Indeed, if right-handed 
sneutrinos get a vev  $\langle{\tilde \nu}_R \rangle=v\leq M_s$, the right-handed neutrinos get a mixing mass term of order $\lesssim g_{Q_\nu}M_s\simeq 0.1-500$ eV with the $U(1)$ gaugino. Thus the gaugino may be  present as a `sterile neutrino' which could play a role in neutrino oscillation physics.

\end{itemize}

The above is just a partial list of phenomenological implications. Thus one expects the presence of gonion copies of the SM quarks and leptons which would be  below the string scale $M_s\simeq 700$ TeV. Some gonions, which would look like vector-like partners of quarks and leptons,  could have masses around a few TeV. In addition, once SUSY is broken, the SUSY partners of the SM would be expected below the gonion scale. Thus there could be a plethora of new particles above LHC scales, possibly within FCC reach, if some of the particles remain light as in the `mini-split' scenario of \cite{Arvanitaki_2013,arkanihamed2012simply,Hall_2013} in which a large SUSY breaking scale up to $\sim 10^3$ TeV is considered. Also, $\nu_R$ gonions or lightest KK particles could be candidates for dark matter. We leave these possible phenomenological implications to future research.

\subsubsection*{Challenges}

 An important question is that of baryon stability. Indeed with a fundamental scale as low as $M_s\simeq 10^6$ GeV,  there is a danger of too fast baryon decay:  In a SUSY setting the most dangerous operators are those of dimension 4 which violate R-parity in the superpotential, $(Q_LD_RL),(U_RD_RD_R),(LLE_R)$. Those are all forbidden 
 by the $U(1)_\nu$ symmetry, as well as the dimension 5 operators $(Q_LQ_LQ_LL)$ and $(U_RU_RD_RE_R)$. One can also check that the standard $6d$ operators of baryon decay appearing in GUT's are also forbidden. It would be interesting to check whether this suppression may be sufficient to ensure baryon stability in a setting like this, once one includes 
 operators of even higher dimensions.
 
 There is another challenge present in this particular setting, which is that `$\mu$-term' bilinears $\mu H_uH_d$ are also forbidden by the  $U(1)_\nu$ symmetry. Although not as crucial as baryon stability, the absence of such a term 
 may be problematic for the Higgs and Higgsino sectors of SUSY theories, since then $H_u$ and $H_d$ may not align when getting a vev. 
 In fact  a $\mu$-term may in principle be induced at the non-perturbative level  from charged instantons, along the lines of  refs.
\cite{Ibanez:2006da,Blumenhagen:2006xt,Cvetic:2007ku,Ibanez:2007rs,Cvetic:2008hi,Ibanez:2008my,Blumenhagen:2009qh,Anastasopoulos:2010hu}. In the present case the existence of Euclidean instantons with charges under $U(1)_b \times U(1)_c \times U(1)_{\tilde c}$ would be required. Checking whether the required instantons exist 
 would require a fully-fledged string compactification. We do not think that this specific $\mu$-problem is generic for any possible semirealistic brane configuration leading to small neutrino Yukawas.  An example of this is the $SU(5)$ model described in appendix \ref{su5}. There a $\mu$-term is allowed,  not forbidden by any symmetry and still one can obtain suppressed neutrino Yukawa couplings in a way similar to the one considered in the above SM-like configuration.

\section{The cosmological constant and Swampland constraints }\label{s:thecc}

In the previous sections we have not considered any input from Swampland criteria \cite{Vafa:2005ui,Arkani-Hamed:2006emk}. Briefly speaking, the Swampland Programme (see \cite{Brennan:2017rbf,Palti:2019pca,vanBeest:2021lhn,Grana:2021zvf}
for reviews) attempts to identify 
the general patterns that an EFT must obey so that it may be part of a consistent theory of 
Quantum Gravity (QG). One of the best tested Swampland conjectures is the Weak Gravity Conjecture \cite{Arkani-Hamed:2006emk},
which states that in any $U(1)$ theory coupled to QG there must exist a particle of charge $q$
with mass obeying $m\leq \sqrt{2}qgM_{\rm P}$. In the present case the tower of gonions satisfies
this condition. In addition, the Swampland distance conjecture \cite{Ooguri:2006in} (SDC) states that as we move in moduli space at infinite distance a tower of states must become massless. In the setting discussed in this paper,
at large $u\rightarrow \infty $ towers of gonions and KK/winding states appear. 

Other classes of Swampland conditions depend on the structure of the vacua of scalar potentials. These other hypotheses have not been tested at the same level as the WGC or the SDC but their implications may be quite important and their validity should be considered seriously. In this section we want to briefly describe some implications of this class
of constraints as applied to our Dirac neutrino setting.

The AdS instability conjecture \cite{Ooguri:2016pdq,Freivogel:2016qwc}
states that there are no consistent  stable (non-SUSY) Anti-de-Sitter (AdS) 
vacua in string theory. This does not sound very useful since the observed universe seems to have
positive cosmological constant rather than negative. Still it has been 
observed \cite{Arkani-Hamed:2007ryu} 
that upon compactification of the SM on a circle AdS vacua appear unless 
the lightest neutrino mass 1) is Dirac and 2)  is sufficiently light, in particular (see 
\cite{Ibanez:2017kvh,Gonzalo:2021zsp,Hamada:2017yji} for details) one finds that it must satisfy 
\beq
m_{\nu, 1} \, \leq a\, \Lambda_{{\rm cc}}^{1/4} \, ,
\label{estupenda}
\eeq
where $a$ is a computable number of order one and $\Lambda_{{\rm cc}}$ is the cosmological constant.
The origin of this bound is that the Casimir energy in 3d has a positive contribution from this lightest neutrino which is enough to avoid AdS vacua to develop, if the above constraint is 
obeyed. This works for Dirac neutrinos, which carry 4 degrees of freedom, but not for Majorana 
which only have 2, not enough to overcome the 4 bosonic degrees of freedom from photon and graviton.
This nicely fits with the scheme we are exploring in this paper where neutrinos have tiny Dirac masses. This nice connection between neutrino masses and the cosmological constant is lost if neutrinos are Majorana.

The neutrino mass for the first generation is given by $m_{\nu, 1}=Y_{\nu,1}|\langle{ H_u}\rangle|$. In our scheme, the Yukawa coupling of the first generation may be written as
$Y_{\nu,1}=g_{\nu}\delta^1$.  One can then write
\beq
g_{\nu} \delta^1 |{ H_u}| \, \lesssim \, \Lambda_{{\rm cc}}^{1/4} \, ,
\eeq
so one gets the bound on the gauge coupling
\beq
g_\nu \, \lesssim \, \frac {\Lambda_{{\rm cc}}^{1/4}}{\delta^1 |{ H_u}|} \, .
\eeq
For fixed  EW scale and cosmological constant, this implies a very small value for $g_\nu$ and therefore a very light tower of gonion and KK-like states, that result in two large dimensions. Moreover, because all neutrino masses are suppressed by $g_\nu$, this explains  why Dirac neutrinos are so light. This is required for the lightest neutrino to be sufficiently light to obey the Swampland bound (\ref{estupenda}). One can also turn around the argument and write
\beq
|{H_u}|\ \lesssim \ \frac {\Lambda_{{\rm cc}}^{1/4}}{\delta^1 g_{\nu}} \ .
\eeq
This shows that the EW scale is bounded and stable and related to the 
value of the c.c. Plugging the values 
$\Lambda_{{\rm cc}}^{1/4}\simeq 10^{-12}$ GeV, $g_{\nu}\simeq 10^{-12}$ and e.g. $\delta^1\simeq 10^{-2}$
one would recover $|{ H_u}|\lesssim 10^2$ GeV. In this sense this provides a solution to the
EW instability (hierarchy) problem, as already pointed out in
refs.\cite{Ibanez:2017oqr,Ibanez:2017kvh,Gonzalo:2018dxi,Castellano:2023qhp}.

To sum up, from a Swampland viewpoint, a Dirac character for the neutrino is required to avoid the Swampland AdS instability conjecture. Additionally, the bound \eqref{estupenda} relates the neutrino mass scale to the cosmological constant and, combined with our scheme, explains why two large dimensions are required to have the lightest neutrino sufficiently light (see also \cite{Castellano:2023qhp}).

\section{Summary and outlook}
\label{s:conclusions}

In this paper we have studied under which conditions neutrino Yukawa couplings may become asymptotically small
	in the moduli space of SM-like string compactifications. This would give rise to tiny Dirac masses for neutrinos once the
	Higgs gets a vev, which may be compatible with neutrino oscillation data. To work out this study we have used as a laboratory type IIA CY orientifolds with D6-brane configurations yielding the chiral content of the SM at their intersections. We believe however that, given the extended network of dualities among 4d, ${\cal N}=1$ string vacua, our results are
	expected to be quite general. In order to perform this analysis we have made use of the recent results of \cite{Casas:2024ttx}, in which the general behaviour of Yukawa couplings
	at infinite distance in moduli space was explored within the context of type IIA CY orientifolds.  
	
	While we want to consider limits along which the neutrino Yukawas are rendered tiny, at the same time we have to ensure that the  SM gauge coupling constants as well as the quark and charged lepton Yukawas do not get a strong suppression along them. This condition turns out to be extremely constraining. Essentially it requires that a large complex structure direction (say $u\rightarrow \infty$) should be
	taken so that two extra compact dimensions become large with a mass scale $\sim  M_{\rm P}/u$. At the same time a tower of 	$\nu_R$-like states becomes also light around the same scale. The neutrino Yukawa couplings are of order $Y_\nu \simeq u^{-1/2}$,
	and one also has $Y_\nu\simeq g_\nu$, where $g_\nu$ is the gauge coupling of the $U(1)$ gauge symmetry coupling to right-handed
	neutrinos, which is thus very weak.  Fixing the neutrino Yukawa couplings at values consistent with neutrino data gives a value $g_\nu \simeq  7\times 10^{-13}$ so that one finds a lowered string scale $M_s\simeq g_\nu M_{\rm P}\simeq 700$ TeV, and the $\nu_R$ tower and extra large dimensions at a scale $\simeq 500$ eV.  
 If one allows for a lack of universality in the $\nu_R$ gonion towers, these scales may be 
 substantially lowered down to  $10$ TeV and $0.1$ eV respectively. Thus our scenario a priori does not exclude a string scale not far away from LHC limis ($\sim 8$ TeV).

	We have also examined the case of a single large dimension within the context of type IIA CY orientifolds with SM content. We find that one large extra dimension cannot explain the tiny value of the Dirac neutrino. Moreover, regardless of the question of neutrino masses, it could be compatible  with the existence of a realistic Yukawa coupling structure \textit{only if} there is no light SM gonion tower along the large dimension. On the contrary, a light gonion tower along it would produce very suppressed Standard Model couplings. Further, we have also considered the possibility that four or six dimensions become large, and when applied to scenarios with tiny Dirac neutrino masses, result in a quantum gravity cutoff that is too low and already experimentally ruled out.
	
	To sum up, we find that tiny Dirac neutrino masses may be obtained in string theory, by imposing a lowering of the string scale 
	to values $M_s\lesssim 10-700$ TeV and the existence of two large dimensions at a scale with a factor $\sim 10^4$ heavier than that of 
	neutrino masses. This is at variance to what happens in the case of Majorana  neutrino  masses which require a large string scale
	$M_s\gtrsim 10^{14}$ GeV. It is also interesting to remark that, as emphasized in section  \ref{s:thecc},	the AdS instability conjecture applied to the 3d SM implies that neutrinos should be Dirac. The latter arguments 
	also give an explanation for the proximity of the neutrino and the cosmological constant scales. 
	
	The case of Dirac neutrino masses may potentially lead to phenomenology much richer than the Majorana case. Indeed, although the
	most conservative case yields a value for the string scale $M_s\simeq 700$ TeV, if there is no universality among the intersection angles of the
	three families of neutrinos, that scale could be as low as $M_s\simeq 10$ TeV, a region which is already being tested 
	by LHC and astrophysical data.  There is also the 
	possibility of a light vector boson with tiny gauge coupling $\sim 10^{-12}- 10^{-14}$ and a mass $M_{V_\nu}\gtrsim 10^{-3}-0.1$ eV which, for the lightest limit, is also	constrained by a combination of fifth-forces and astrophysical limits. It would be important to make a systematic analysis of these and other phenomenological consequences of Dirac neutrinos in the context of string theory.  Furthermore, it would also be interesting to search for complete, tadpole free examples of type II  orientifolds with SM-like spectra with the required conditions, as well as realisations in other corners of the Landscape of SM-like string constructions. In the meantime, it is remarkable how knowledge of the Dirac or Majorana character of neutrinos can give us so much information about the structure that realistic string SM-like vacua should have.

\bigskip

\vspace*{.45cm}

\centerline{\bf \large Acknowledgements}

\vspace*{.5cm}

We  thank Alberto Castellano, Jos\'e Luis Hernando, \'Alvaro Herr\' aez,  Luca Melotti and  \'Angel Uranga for discussions.  This work is supported through the grants CEX2020-001007-S and PID2021-123017NB-I00, funded by MCIN/AEI/10.13039/501100011033 and by ERDF A way of making Europe. G.F.C. is supported by the grant PRE2021-097279 funded by MCIN/AEI/ 10.13039/501100011033 and by ESF+. 


\appendix

\section{The case of a single large dimension and Yukawa couplings}
\label{onedimension}

We have seen that starting with a 4D $N=1$ SUSY vacuum, obtaining small neutrino Yukawa couplings
naturally leads to two decompactified dimensions. It is interesting to see however whether the
case with a single large dimension and a small neutrino Yukawa coupling is also viable.  
  
  Let us consider as a guideline the ${\bf Z}_2\times {\bf Z}_2$ toy model  with gauge group
  $U(4)\times Sp(2)_R\times Sp(2)_L$ described in the main text.
  Now, we know that in the i-th torus we have KK and winding states, given by
  \be
m_{{\rm KK},i}\, =\, \frac {2M_s}{R_{2i-1}}\, =\, \frac {M_{\rm P}}{2A_i^{1/2}\sqrt{su^{(i)}}} \, , \qquad 
	 m_{{\rm w},i}\, =\, \frac {1}{2}R_{2i}M_s\, =\, \frac {A_i^{1/2}M_{\rm P}}{2\sqrt{su^{(i)}}}  \,  .
	 \label{masillasA}
  \ee
	 We want to take the limit in which $R_{2i-1} \sim u  \rightarrow  \infty$ for some $i$,  (implying $s,u^{(i)}\rightarrow u$, with $e^{-\phi}$ invariant) 
	   while keeping $R_{2i}$ constant in string units, so we are left with only one large dimension. Since ${\rm Im}T_i=t^{(i)}=t=A_i$, we need to take the limit:
\be
s, u^{(i)}\ \sim 	  \ u \ ;\ t^{(i)}  \sim \ u \ ;\ u\longrightarrow \ \infty \ .
\ee
One then has for the KK and  winding scales along the i-th plane  the following scaling
\beq 
m_{{\rm KK},i} \ \simeq \ \frac {M_{\rm P}}{u^{3/2} } \ ;\ m_{{\rm w},i}\ \ \simeq \ M_s \ \simeq \ \frac {M_{\rm P}}{u^{1/2}}.
\eeq
In order to obtain some tiny Yukawa coupling we need to take the limit along a complex plane 
in which that limit implies a light tower of gonions. In the toy example that means either $i=2$ or $i=3$.
So let us consider for definiteness $i=2$.

There are several quantities that only depend on the complex structure. In particular the gauge kinetic function and the
FI-terms.  Also the gonion masses. Thus one has:
\be
g^2\equiv g_a^2\ \simeq \ g_R^2 \ \simeq \frac {1}{u}\ ;\  m_{gon}\ \simeq \ \frac {M_{\rm P}}{u} \ .
\ee
 Recalling the expression for Yukawa couplings
  \be
	 Y_{ijk}\, =\, e^{\phi_4/2}  {\rm Vol}_X^{1/4} W_{ijk} \Theta_{ijk}^{1/4} \, ,\label{yukapp}
	 \ee
and writing  ${\rm Vol}_X = {\rm Vol}_4\times A_2$, in the studied limit we get
\be
	 Y_{{F_R}F_LH} \ \simeq \    t^{1/4}\left(\frac {m_{{\rm gon},2}}{M_{\rm P}}\right)^{1/2}\ \simeq
	 \ \frac {t^{1/4}}{u^{1/2}} \ \simeq \  g^{1/2}
	 \ee
This is at variance with the 2-large dimension case in which $Y\sim g$. 
Also at variance is the scale of the large dimension  $R_3$ which in this case is given by
\be
m_{\rm KK}\ \simeq \ \frac {M_{\rm P}}{t^{1/2}u}\ \simeq  \frac {M_{\rm P}}{u^{3/2}} \ \simeq \ g^3 M_{\rm P},
\ee
whereas in the case of 2 large dimensions is $m_{\rm KK}\sim g^2M_{\rm P}$.  Note that in the case of 2 large dimensions
 ${\rm KK}$, winding and gonion masses are of the same order. In the case of 1 large dimension the gonions 
 masses are  $g^{-1}$ times larger.
The string scale in this case is still $M_s=e^{\phi_4}M_{\rm P}\simeq  M_{\rm P}/(u^{1/2})$. 
It corresponds to the species scale since:
\be
\Lambda \ \simeq (m_{\rm KK})^{1/3}M_{\rm P}^{2/3} \ \simeq gM_{\rm P} .
\ee 

\subsection{Application to a small Yukawa neutrino setting}

One can translate this structure to e.g. the brane setting with the SM. Now the point is that
imposing that the Yukawa has the required size one fixes the value of $g_\nu\simeq 1/u^{1/2}$:
\be
Y_\nu \, \simeq \, g_\nu^{1/2}\, \simeq \, 7\times 10^{-13}\, \implies \,  g_\nu \, \simeq \, 10^{-25}\, .
\ee
This value for $g_\nu$ is much smaller than in the case of 2 large dimensions which has $g_\nu\simeq 10^{-12}$.
But such small value for $g_\nu$ implies a KK scale for the large dimension $m_{\rm KK}\simeq g_{\nu}^3M_{\rm P}\simeq 10^{-57}$ GeV,
and a ridiculously small species scale $\Lambda\simeq m_{\rm KK}^{1/3}M_{\rm P}^{2/3}\simeq gM_{\rm P}\simeq 100$ eV.
Thus a single dimension does not seem to work if you want to obtain Dirac neutrino masses from
tiny Yukawa couplings.

\subsection{One large dimension and the  `Dark Dimension' scenario}

We have just seen that a single dimension does not work for the purpose 
of understanding the smallness of Dirac neutrino masses.
Still we may asume that there exists a single large dimension because of other motivations.
This is the case of the dark dimension scenario \cite{Montero:2022prj,Gonzalo:2022jac,Obied:2023clp,Anchordoqui:2023oqm}.
In this scenario one imposes that there is a tower with $m_{\rm KK}\simeq  V^{1/4} \simeq 10^{-12}$ GeV,
related to the cosmological constant.
Note that we saw that  in this limit
\be
m_{\rm KK} \ \simeq \ \frac {M_{\rm P}}{u^{3/2} }\ ,
\ee
so that in order to get $m_{\rm KK}\simeq 10^{-3}$ eV, one needs $u\simeq 10^{20}$ and $\Lambda \simeq 10^8$ GeV.

Let us now explore whether one can have a consistent structure for all the SM Yukawa couplings, although 
giving up on having a tiny Dirac neutrino mass. It is useful to study two cases separately
depending on whether {\it i)} There is no tower of light gonions along the extra large dimension or {\it ii)}  there is a tower of gonions.  By gonion tower, we are referring here to intersections at which SM fields live.

{\it i) One large dimension without light gonion tower}

An example of this will be taking the limit $u^{(1)}\rightarrow \infty$ in the first complex plane 
in the toy model. It is easy to see that in this model in that limit the
angles at the intersection  of the matter fields $F_R,F_L,H$ remain finite, and hence there is no tower of light gonions (there are gonions at the $(aa^*)$ intersections but the corresponding zero modes do not have Yukawa coupings to the rest). Let us consider then the Yukawa coupling $(F_RF_LH)$. It will have a structure
\be
Y_{{\cal N}= 1} \ \simeq \ t^{1/4}   e^{\phi_4/2} \left( \frac {m_{{\rm gon},1}}{M_s}\right)^{1/2}\left( \frac {m_{{\rm gon},2}}{M_s}\right)^{1/2}\left( \frac {m_{{\rm gon},3}}{M_s}\right)^{1/2} \ .
\label{yukuno}
\ee
Now, all gonions in all three planes  will have masses of order $M_s$, so that the Yukawa couplings will be of order $Y\simeq  t^{1/4}e^{\phi_4/2}\simeq 1$.
Going to a more realistic SM setting as described in the main text, 
this is ok  phenomenologically for the heaviest quark-lepton generation. Smaller Yukawa couplins for the first generations
could come both from $(m_{\rm gon}/M_s)^{1/2}$ powers or perhaps the holomorphic superpotential.\footnote{It must be said that there is a point in which the structure of the toy model departures from a
realistic SM configuration. Indeed, there are gonions at the $(aa^*)$ intersections in the toy model. This fact comes along with a contribution to the $U(4)$ gauge coupling which decreases like $\sim 1/u^{1/2}$.
This would not happen in a SM configuration in which $(aa^*)$ intersections are absent, so there is no need for
any SM gauge coupling to become too small.}

So as a summary: a single large dimension along a direction which does not feature light gonions 
associated to the SM is compatible with the existence of a realistic structure of Yukawa couplings.

{\it ii) One large dimension with light gonion towers}

Consider now the case in which the large dimension goes along a complex plane in which a light tower of gonions 
develop. This is for example the case of the second plane in the toy model with $s,t^{(2)},u^{(2)}\sim u$,
$u\rightarrow \infty$. 
The Yukawa couplings with a `${\cal N}=1$ structure' would look like in eq.(\ref{yukuno}) above.
However now the Yukawa is further supressed by a factor   $  ({m_{{\rm gon},2}}/ {M_s} )^{1/2} \simeq 1/u^{1/4}\simeq 10^{-5}$.
There are now two subcases, depending on whether the tower of gonions corresponds to 
right-handed neutrinos or to the rest of the SM particles:

{\it 1) The gonion towers correspond to right-handed neutrinos }.
In this case there will be a tower
of $\nu_R$-like gonions at a scale $m_{gon,\nu}\simeq 10^{-10}M_s\simeq 10$ MeV.  Neutrinos will get a Yukawa of order
$Y_\nu\simeq 10^{-5} $. One does not expect a large  instanton generated Majorana mass for the right-handed neutrinos, 
since they would be exponentially suppressed like $e^{-t}$. Thus a usual see-saw would not be available and  perhaps 
some other mechanism involving the massive gonions could be at work.  Asuming this is possible, indeed 
a large single dimension along  a direction involving $\nu_R$ gonions could be consistent.

{\it 2) The gonion towers correspond to  gonions of SM particles}

In this case the above Yukawa would be 
too small to be consistent with the heaviest quark/lepton generations. So such type of Yukawa couplings would not be
enough to describe the observed SM spectrum. Still, one can consider Yukawa couplings with a 
`${\cal N}=2$' type of structure.
In this case the Yukawa couplings will have the form
\be
Y_{{\cal N}= 2} \, \simeq \, t^{1/4} g_{\rm SM} \left( \frac {m_{{\rm gon},1}}{M_s}\right)^{1/2}\left( \frac {m_{{\rm gon},3}}{M_s}\right)^{1/2}  \, .
\ee
Here the factor $g_{\rm SM}$ stands for a generic SM gauge coupling and comes from the prefactor $h_{22}^{-1/2}$
in eq.(3.7) in ref.\cite{Casas:2024ttx}.
Now Yukawa couplings as large as $Y\sim 1$, as required by the top quark, may be easily obtained. In fact the problem is the contrary since there is the prefactor $t^{1/4}\simeq u^{1/4}\sim 10^{5}$ which makes the Yukawa too large unless it is compensated by the gonion factors in the right. So, assuming SUSY, Yukawa couplings as large as one, 
but not larger, are possible if
\be
m_{{\rm gon},2} \ \simeq m_{{\rm gon},3} \lesssim \ \frac {1}{u^{1/4}} M_s \ \simeq \ 10^{-5} M_s,
\ee
where we have taken conservatively $g_{\rm SM}\sim 1$
(if SUSY is not assumed gonions in these sectors  could  be as light as $\simeq 10^{-10}M_s$, which would be ruled out).
With $M_s\simeq 10^8$ GeV, this means that up to 4  new extra large dimensions should develop  at a scale $\lesssim 1$ TeV or below, which is already ruled out experimentally. Furthermore the SM gauge couplings would be too small due to the contribution to the
running of gauge couplings from the towers of SM gonions.

So as a second summary: {a single large dimension along a direction which  features light $\nu_R$ gonions 
is consistent. However if the gonions are 
associated to the SM particles, it is not viable}

\subsubsection*{Some difficulties with more than two large dimensions}

One may also think of considering models with more than two large dimensions.
One can see however that if we were able to build a model leading to tiny neutrino 
Yukawa couplings with more than two large dimensions, the string scale would be lowered
much below the LHC threshold. We will just give a heuristic idea which goes
as follows. The species scale $\Lambda_{\rm QG}$ (which is our setup is the string scale) generated by KK-towers of characteristic mass $m_t$ is computed in Planck units as (see e.g. \cite{Castellano:2021mmx})
\be
\Lambda_{\rm QG} \, \simeq \, (m_t)^{\frac{p}{D-2+p}} \, ,
\ee
with $D$ the number of dimensions and $p$ coincides with the number of decompactified 
dimensions (or $p=2$ in the case of gonion towers).  In the main text we have studied 
the case $D=4$, $p=2$ which leads to $\Lambda_{\rm QG} \simeq (m_tM_{{\rm P}})^{1/2}$. Now, let us assume 
that instead, we have $p=4,6$ and that all towers scale in the same way, $m_{\rm KK}\sim m_{{\rm w}}\sim  1/u$. In this  case, by the same assumptions used below \eqref{Yijnu}, it will still be true that
$Y_{\nu} \simeq u^{-1/2}\simeq g_\nu$. So we will have the species scale of the order of
\be
\Lambda_{\rm QG} \, \simeq \, u^{- \frac{p}{d-2+p}} M_{\rm P} \, \simeq \, Y_{\nu}^{\frac{2p}{2+p}} M_{\rm P} \, .
\ee
If we now set the value of the Yukawa to $Y_\nu\simeq 7\times 10^{-13}$, for $p=4,6$ 
one obtains a species scales of order $100$$,\ 1$ GeV  respectively, which are obviously ruled 
out experimentally.

\section{\texorpdfstring{$SU(5)$}{Lg}  quiver with a small neutrino Yukawa}
\label{su5}  

Here we consider the generation of a small neutrino Yukawa coupling in a GUT-like brane configuration with three brane stacks and gauge group $U(5)\times U(1)\times U(1)$. This example was considered in \cite{Anastasopoulos:2010hu}, where it is shown how that structure may arise in CFT Gepner type II orientifolds along the lines of 
\cite{Dijkstra:2004ym,Anastasopoulos:2006da,Kiritsis:2008ry,Kiritsis:2009sf}. 
The spectrum at the non-vanishing intersections is shown in table \ref{GUTSU(5)1}.
\begin{table}[h!!]\begin{center}
\renewcommand{\arraystretch}{1.00}
\begin{tabular}{|c|c|c|c|c|c|}
\hline
Intersection & $SU(5)$  &  $U(1)_5$ & $U(1)_b$  & $U(1)_c$ & Inter. \\
\hline\hline
$aa^*$ &  $10$ &   $2$  &  0  & 0 & $+3$ \\
\hline
$ab$ &  ${\bar 5}$ &  $-1$ & 1 & 0 & $-3$\\
\hline
$ac$ &  $5_{H_u}+{\bar 5}_{H_d}$ &  $\pm 1$ & $0$ & $\mp 1$ & $\pm 1$\\
\hline
$bc$ &  $1 $ &  $0$ & $-1$ & $1$ & $-3$ \\
\hline
\end{tabular}
\caption{ Configuration of $D6$-brane intersections in a 3-stack orientifold  $SU(5)$ model.
}
\label{GUTSU(5)1}
\end{center}
\end{table}
In this model the right-handed neutrino is identified with the singlet at the intersection $bc$. The neutrino Yukawa coupling
\beq 
Y_{\nu,ij}({\bar 5}^i5_{H_u})\nu_R^j \ 
\eeq
is perturbatively allowed from the intersection of the $a,b$ and $c$ stacks. The gauge group felt by the $\nu_R$'s is $Q_\nu=Q_b-Q_c$. The generation of a small Yukawa goes along the lines described in the main text for the SM-like model with five stacks and also the toy model. The angle at the $\nu_R$ intersections decreases like $\theta\sim 1/u$, with $u$ the imaginary part of a complex structure field $U$.
Just like in those examples, one has for the gauge coupling that
\beq
\re f_{Q_\nu}\, \simeq u \, \longrightarrow \, g_\nu \, \sim \, \frac {1}{u^{1/2}} \, .
\eeq
Again, the $\nu_R$ gonion mass scale is $m_{{\rm gon},\nu}\sim M_{\rm P}/u\simeq g_\nu^2M_{\rm P}$,
and the string scale $M_s\sim e^{\phi_4}M_{\rm P}\simeq g_\nu M_{\rm P}$. For the Yukawa coupling at large $u$ we have
\beq
Y_{\nu,ij} \simeq  e^{-\phi_4} \left(\frac {m_{{\rm gon},\nu}}{M_{\rm P}}\right)^{1/2}
\left(\frac {m_{{\rm gon},{\bar 5}}}{M_{\rm P}}\right)^{1/2}
\left(\frac {m_{{\rm gon}, 5_{H_u}}}{M_{\rm P}}\right)^{1/2}
\simeq 
\left(\frac {m_{{\rm gon},\nu}}{M_{\rm P}}\right)^{1/2} (\theta_{\bar 5}\theta_{5_{H_u}})^{1/4} \ ,
\eeq
so that for the heaviest neutrino, one arrives at
\beq
Y_{\nu,3}\, \simeq \, \left(\frac {m_{{\rm gon},\nu}}{M_{\rm P}}\right)^{1/2} \simeq \, g_\nu .
\eeq
Setting $g_\nu \simeq 7\times 10^{-13}$ one reproduces a neutrino mass 
consistent with neutrino oscillation results.

As pointed out in \cite{Anastasopoulos:2010hu} this brane configuration only allows at the perturbative level for the discussed neutrino Yukawa couplings and a 
$\mu$-term, a bilinear ${\bar 5}_{H_d}5_{H_u}$. The remaining Yukawa couplings $10\times 10\times 5_{H_u}$ and $10\times {\bar 5}\times {\bar 5}_{H_d}$ 
must appear through instanton effects. In particular Euclidean 2-branes
$E2_1$ and $E2_2$ with charges $(-5,0,1)$ and $(0,2,-2)$ respectively could give rise to this Yukawa terms. 
To check whether the needed instantons exist or not would require
a complete study of non-perturbative effects in a complete orientifold compactification. In this sense, an F-theory avatar of this kind of vacua would be interesting.

\section{ Anomalies and massive \texorpdfstring{$U(1)$'s}{Lg}  in the 5-stack SM-like quiver}
\label{anomalias}  
 
Here we review the SM brane configuration considered in section \ref{s: SM intersecting}, together with the conditions imposed by the absence of field theory anomalies. The total gauge group is given by $SU(3)\times SU(2)\times U(1)$ and is extended by two additional $U(1)$ Abelian groups. The spectrum of charges of the model is summarised in table \ref{tablaSM-2}, corresponding to the following intersection numbers $I_{\a\b}$ 
\bea
& I_{ab}= 3,\quad I_{a\Tilde{c}}=-3,\quad I_{ac}=-3,\quad I_{bd}=-3,\label{intersections}\\
& I_{\Tilde{c}d}=3,\quad I_{bc}=K,\quad I_{b \Tilde{c}}=\Tilde{K},\quad I_{cd}=3,\notag
\eea
where $K$ and $\Tilde{K}$ are the number of Higgs doublets that will be fixed by anomaly cancellation. First, the cancellation of $U(N_i)$ anomalies requires 
\be
\sum_j I_{ij}N_j =0,
\ee
with $N_a =3$, $N_b =2$, $N_c = N_{\tilde{c}} = N_d = 1$. Threfore we obtain
\bea
&\sum_j I_{aj}N_j =\sum_j I_{bj}N_j =\sum_j I_{dj}N_j=0 \\
& \sum_j I_{cj}N_j =12 -2 K=0,\quad \sum_j I_{\Tilde{c}j}N_j =12 -2 \Tilde{K}=0
\eea
which fixes the number of Higgs doublets to $K=\Tilde{K}=6$. Moreover, the triangle mixed anomalies of the $U(1)$'s with the non-Abelian $SU(N_j)$ groups are given by
\be
\mathcal{A}_{ij}=\frac{1}{2}I_{ij} N_i.
\ee
For the $SU(3)$ group they read
\bea
& \mathcal{A}_{ba}= \frac{1}{2}I_{ba} = -3,\quad \mathcal{A}_{ca} = \frac{3}{2},\quad\mathcal{A}_{\Tilde{c}a}=\frac{3}{2},\quad \mathcal{A}_{d a}= 0,
\eea
while for the $SU(2)$ 
\bea
& \mathcal{A}_{ab} = \frac{9}{2},\quad \mathcal{A}_{cb} = -3,\quad\mathcal{A}_{\Tilde{c}b}=-3,\quad \mathcal{A}_{d b}= \frac{3}{2}.
\eea
As can be easily checked, the hypercharge, defined as $Y=\frac{2}{3} Q_a+\frac{1}{2} Q_b+Q_c$ is anomaly-free. Likewise, the model has two extra anomaly-free directions, namely 
\be
\Tilde{Q}=  Q_c - Q_{\Tilde{c}}\, ,\quad Q_{B-L}= -\frac{1}{3}Q_a + Q_d\, . 
\label{eq: extraU}
\ee
As pointed out in the main text, the group $U(1)_{\nu}$ under which the neutrinos are charged is anomalous and reads
\be
Q_{\nu} = Q_{\tilde{c}}-Q_{d}.
\ee
For a realistic construction, every Abelian group must acquire a mass besides $U(1)_Y$. Anomalous Abelian groups already satisfy this, since they acquire mass by a Green--Schwarz mechanism \cite{Aldazabal:2000dg,Ibanez:2001nd}. On the contrary, anomaly-free groups could potentially be massless. However, as pointed out in section \ref{s: SM intersecting}, as long as the branes $a$, $b$, and  $c$ belong to a localised sector of the compactification, and branes $\tilde{c}$ and $b$ are sensitive to the bulk moduli, we expect all those additional anomaly-free $U(1)$'s to be massive.

\bibliographystyle{JHEP2015}
\bibliography{bibliography}

\end{document}